\documentclass[12pt]{iopart}

\usepackage{bbm}
\usepackage{graphicx}
\usepackage{array}
\usepackage{bm}

\renewcommand{\vec}[1]{\mathbf{#1}}
\newcommand{\vect}[1]{\boldsymbol{\mathbf{#1}}}\renewcommand{\imath}[0]{\mathsf{i}}
\makeatletter
\newcommand{\undersim}[1]{\mathrel{\mathpalette\@undersim{#1}}}
\newcommand{\@undersim}[2]{%
  \vcenter{%
    \ialign{%
      ##\cr
      $\m@th#1#2$\cr
      \noalign{\nointerlineskip\kern.2ex}
      $\m@th#1\sim$\cr
      \noalign{\kern-.4ex}
    }%
  }%
}
\newcommand{\gsim}{\undersim{>}}
\renewcommand{\lsim}{\undersim{<}}
\makeatother
\usepackage[normalem]{ulem}
\usepackage{color}

\begin{document}

\title{Transport of a passive scalar in wide channels with surface topography}

\author{J. V. Roggeveen$^1$, H. A. Stone$^1$, C. Kurzthaler$^1$$^2$}
\address{$^1$ Department of Mechanical and Aerospace Engineering, Princeton University, New Jersey 08544, USA}
\address{$^2$ Max Planck Institute for the Physics of Complex Systems, 01187 Dresden, Germany}
\eads{\mailto{jamesvr@princeton.edu,} \mailto{hastone@princeton.edu}, \mailto{ckurzthaler@pks.mpg.de}}

\begin{abstract}
We generalize classical dispersion theory for a passive scalar to derive an asymptotic long-time convection-diffusion equation for a solute suspended in a wide, structured channel and subject to a steady low-Reynolds-number shear flow. Our theory, valid for small roughness amplitudes of the channel, holds for general surface shapes expandable as a Fourier series. We determine an anisotropic dispersion tensor, which depends on the characteristic wavelengths and amplitude of the surface structure. For surfaces whose corrugations are tilted with respect to the applied flow direction, we find that dispersion along the principal direction (i.e., the principal eigenvector of the dispersion tensor) is at an angle to the main flow direction and becomes enhanced relative to classical Taylor dispersion. In contrast, dispersion perpendicular to it can decrease compared to the short-time diffusivity of the particles. Furthermore, for an arbitrary surface shape represented in terms of a Fourier decomposition, we find that each Fourier mode contributes at leading order a linearly-independent correction to the classical Taylor dispersion tensor.
\end{abstract}
\maketitle
\section{Introduction}
Transport processes at small scales play a central role in a number of fields, ranging from biology~\cite{Viallat:2019,Bechinger:2016}, where solutes and blood cells are transported through vascular channels or bacteria spread through soil and tissues, to geophysics~\cite{Garcia:2008}, where water flows through complex porous materials and eroded rock particulates sediment in rivers, to microfluidic applications~\cite{Beebe:2002,Stone:2004,ElAli:2006,Whitesides:2006,Sackmann:2014}, where various biological and synthetic samples are mixed, sorted, or focused while moving through microchannels. These systems often operate at low Reynolds number, where viscous forces dominate over inertia and strong stochastic fluctuations due to Brownian effects dictate the overall transport behavior~\cite{Hofling:2013}. In contrast to dilute, unconfined media, natural environments and microfluidic applications exhibit a variety of confining surfaces with different properties, including deformable and elastic materials~\cite{Rallabandi:2018}, fluctuating boundaries~\cite{Marbach:2018}, and structured topographies~\cite{Kurzthaler:2020}, which result in complex hydrodynamic flows. Understanding the interaction of Brownian suspensions with nearby boundaries and flows provides insight into microbiological processes and enables optimization and design of microfluidic devices~\cite{Stone:2004}.

The study of solute transport in microchannels dates back to Taylor~\cite{Taylor:1953} and Aris~\cite{Aris:1956} in the 1950's. In their seminal work, they predicted an enhancement compared to diffusion alone of the streamwise spreading of particles, as diffusion across the channel allows particles to sample different velocities of the hydrodynamic flows. Since then, numerous theoretical and experimental studies have extended their ideas, treating the transient profile of the solute concentration~\cite{Aminian:2016, taghizadeh:2020,Vilquin:2021}, the effect of a finite particle size~\cite{Adrover:2019}, particle-particle interactions~\cite{Griffiths:2012}, channel corrugations~\cite{Hoagland:1985}, channel wall slip~\cite{Ng:2011}, and absorbing and pulsating channel walls~\cite{Marbach:2019}.
The latter can lead to a reduction of particle dispersion due to an entropic slow-down resulting from channel wall constrictions. Such a slow-down has also been reported for particles diffusing through periodic, corrugated channel geometries~\cite{Mangeat:2017}. On the other hand, channel wall topography may enhance dispersion due to the generation of additional hydrodynamic flows and velocity gradients~\cite{Marbach:2018}.

Beyond dispersion through channel geometries, solute transport in porous materials has received substantial scientific attention dating back to Saffman in the 1960's~\cite{Saffman:1960}. Continuing the work of several decades, in the 1970's de Josselin de Jong noted that dispersion in porous materials is a tensorial quantity~\cite{DEJOSSELINDEJONG:1972}, which encodes the (anisotropic) properties of the medium. In 1980 Brenner developed a systematic theory to predict the second-order anisotropic dispersion tensor and drift of particles spreading through spatially periodic, porous media~\cite{Brenner:1980}, which relies on the details of the flow field within one periodic cell. Brenner's theory has been extended to account for fluctuations in the environment that are of the order of the transport process itself~\cite{Koch:1987}. At the same time, other researchers characterized the tensorial nature of diffusion in random porous media \cite{Koch:1988,Sahimi:1993}. Building on Brenner's pioneering work, the study of self-propelled agents in spatially periodic media has demonstrated that obstacles act as entropic barriers while producing additional velocity gradients, which can simultaneously reduce or enhance dispersion, depending on the local shear rate~\cite{Alonso:2019}. Reminiscent of the transport through fluctuating channels mentioned above, experiments revealed that fluctuations of the obstacles enhance transport of particles diffusing through the porous matrix~\cite{Sarfati:2021}. Going beyond periodic structures, other experiments have demonstrated anomalous dispersal of solutes in random porous media, where complex flow patterns of Newtonian and non-Newtonian fluids emerge within the dead end pores~\cite{Bordoloi:2022,Kumar:2022}.

While these studies provide immediate insights into biological and geophysical transport phenomena, they can also guide the design of new microfluidic devices, which rely on the precise control of transport of particulate suspensions~\cite{Beebe:2002,Stone:2004,ElAli:2006,Whitesides:2006,Sackmann:2014}. For example, the method `deterministic lateral displacement' relies on the use of arrays of posts within a microfluidic channel to efficiently separate biological and synthetic constituents of different sizes~\cite{Huang:2004,Davis:2006,McGrath:2014}. The effect of diffusion within this context has been analyzed both experimentally~\cite{Cerbelli:2012} and theoretically~\cite{Cerbelli:2013} in the realm of Brenner's theory. In particular, it has been shown that the interactions of finite-sized particles with the anisotropic obstacle field can generate long-time anisotropic dispersion~\cite{Cerbelli:2012, Cerbelli:2013}.

Another common approach is the use of tailored surfaces with particular surface topographies to achieve mixing~\cite{Stroock:2002}, sorting~\cite{Choi:2007,Choi:2011,Wang:2014,Tasadduq:2017}, or focusing of particles in  suspension~\cite{Hsu:2008,Qasaimeh:2017}. In addition, recently, it has been found that particles moving past herringbone structures have complex, three-dimensional trajectories due to the particle's interaction with the surface~\cite{Chase:2022, Kurzthaler:2022}, yet the effect of diffusion, which in many contexts is not negligible, on their long-time transport properties has not yet been addressed. While in several studies dispersion in narrow channels has been studied, in many of these microfluidic applications channels are significantly wider than they are tall, requiring a new, fully three-dimensional theoretical approach for the characterization of dispersion, which requires consideration of transport both along and perpendicular to the flow.

Here, we revisit the classical Taylor dispersion theory and extend it for scalar transport in wide, structured channels. By ``structured" we refer to the shape of the channel walls. We present an asymptotic long-time, two-dimensional convection-diffusion equation, in contrast to Taylor's one-dimensional equation for dispersion through narrow channels. In particular, the three-dimensional nature of the surface structures no longer allows for a reduction to either a two-dimensional or axisymmetric description of the flow.  Our theory, valid for small surface amplitudes, provides an analytic prediction for the dispersion matrix and the overall drift as a function of the surface shape. We provide results for different surface structures, ranging from corrugated channel walls, as often used in microfluidic applications, to randomly structured topographies. Finally, we use stochastic simulations to corroborate our theory.

\section{Theoretical Background}
\begin{figure}[tp]
\centering
\includegraphics[width = .5\linewidth]{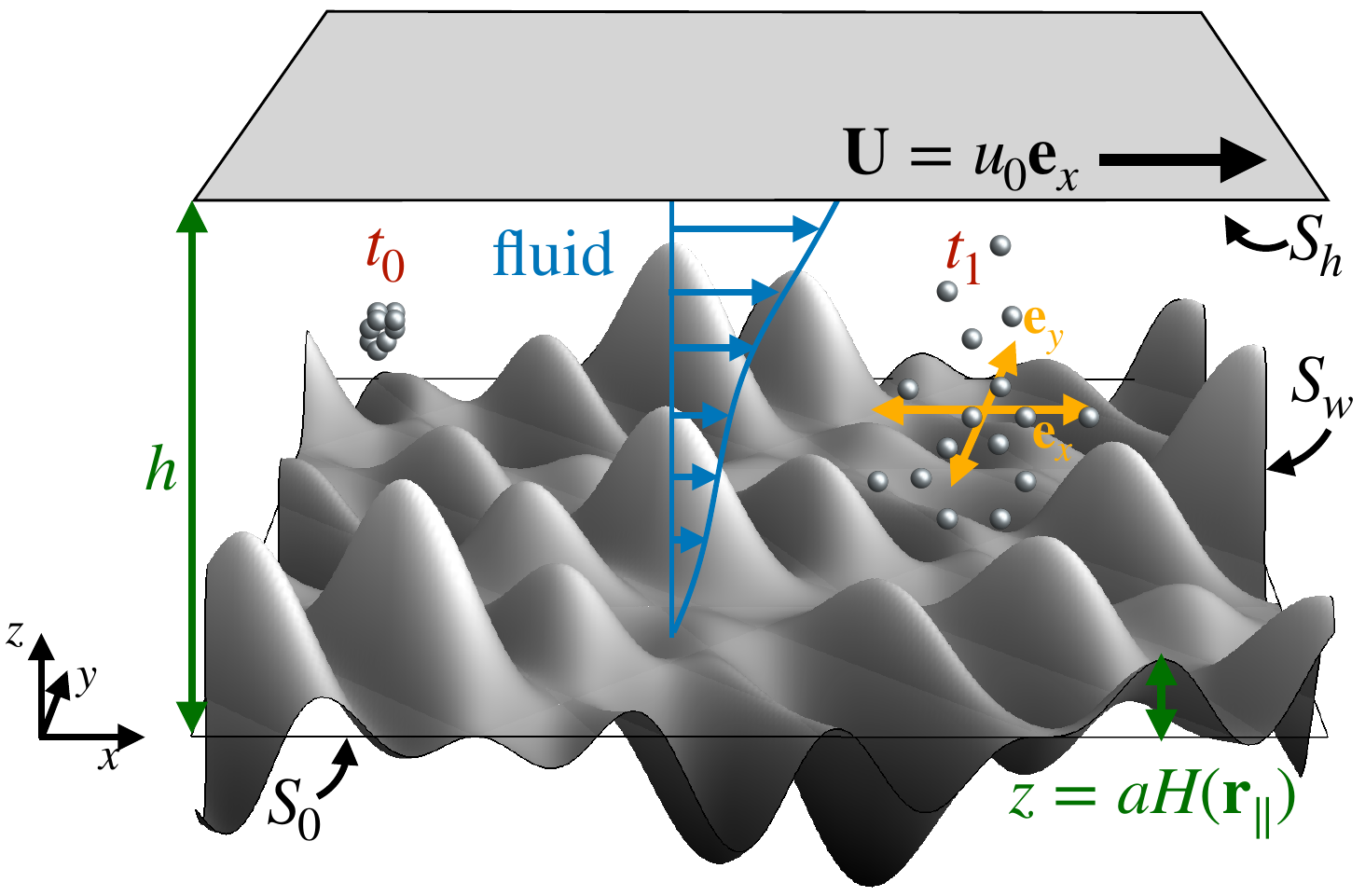}
\caption{Model sketch. Tracer particles are suspended between an upper planar wall, $S_h$, located at $z=h$ and a lower, structured surface, $S_w$. The latter is described by $z=aH(x,y)$, where $H(x,y)$ denotes the shape function and $a$ is the surface amplitude.  Further, $S_0$ corresponds to the reference surface at $z=0$. Sketches of the particle distributions at time $t=t_0$ and a later time $t=t_1$ are shown. The upper wall, $S_h$, moves at a velocity $u_0$ along the $x-$direction. \label{fig:sketch}}
\end{figure}

Consider a Brownian particle at a position $\vec{r}$ at time~$t$. The probability density $c(\vec{r},t)$ of the particle (or, equivalently, the solute concentration) is governed by the Fokker-Planck (or convection-diffusion) equation,
\begin{equation}
    \partial_t c = - \vect{\nabla} \cdot (\vec{u} c) + D_0 \nabla^2 c,
    \label{eq:fokkerplank}
\end{equation}
where $D_0$ is the short-time diffusivity of the particle and $\vec{u} = [u,v,w]^{\rm T}$ is a quasi-steady low-Reynolds-number flow field. The particle is suspended in a channel with a structured lower wall, $S_w$, and a planar upper wall, $S_h$, at $z=h$ (measured from a reference surface $S_0$), see \fref{fig:sketch}. The structured wall $S_w$ is described by the profile $z=aH(\vec{r}_\parallel)$, where $\vec{r}_\parallel=[x,y,0]^{\rm T}$ are the in-plane coordinates, $a$ denotes the characteristic surface amplitude, and $H(\vec{r}_\parallel)$ is the surface shape function. In addition, the upper wall is moving at a speed $u_0$ along the $x-$direction and the lower wall is stationary.

We assume that the channel is not confined along the horizontal directions. At the top and bottom of the channel the probability density obeys a no-flux boundary condition,
\begin{equation}
    \vec{n}\cdot\boldsymbol{\nabla} c(\vec{r},t) = 0 \qquad {\rm on}\; S_h \;{\rm and}\; S_w, \label{eq:no_flux}
\end{equation}
where $\vec{n}$ denotes the vector normal to the upper planar surface $S_h$  and the lower corrugated surface $S_w$, respectively.
We are interested in deriving the behavior of the height-averaged probability density
\begin{equation}
    C(\vec{r}_\parallel,t) = \frac{1}{h-aH(\vec{r}_\parallel)} \int_{aH(\vec{r}_\parallel)}^h c(\vec{r},t)\,\rmd z \equiv\left\langle c(\vec{r},t)\right\rangle,
\end{equation}
where $\langle \cdot \rangle$ represents the height-averaging. Unlike in traditional Taylor-dispersion theory, where averaging over the cross-section of the channel leads to an effective one dimensional transport equation, we retain two horizontal spatial dimensions. We decompose the full probability density as
\begin{equation}
    c(\vec{r},t) = C(\vec{r}_\parallel,t) + \tilde{c}(\vec{r},t),
    \label{eq:probDensity}
\end{equation}
where $\tilde{c}$ is a perturbation term that accounts for the variations in the probability density along the $z$-direction.

Inserting (\ref{eq:probDensity}) into (\ref{eq:fokkerplank}) and defining the in-plane gradient $\nabla_\parallel=[\partial_x,\partial_y, 0]^{\rm T}$ we obtain
\begin{equation}
    \partial_t (C + \tilde{c}) = - \vect{\nabla}\cdot(\vec{u}(C+\tilde{c}))+D_0 (\nabla_\parallel^2 C + \nabla^2 \tilde{c}).
    \label{eq:fullCeqn}
\end{equation}
Height averaging (\ref{eq:fullCeqn}) and recalling the no-flux boundary condition (\ref{eq:no_flux}), we find
\begin{equation}
    \eqalign{
    \partial_t C &= -\langle \vect{\nabla} \cdot(\vec{u}(C+\tilde{c}))\rangle + D_0 \nabla_\parallel^2 C,\label{eq:Ceqn}\\
    &= -\vect{\nabla}_\parallel \cdot(\langle \vec{u}\rangle C) - \langle \partial_z w\rangle C
    - \langle \vect{\nabla}\cdot (\vec{u}\tilde{c})\rangle + D_0 \nabla^2_\parallel C, }
\end{equation}
where by definition $\langle \tilde{c}\rangle =0$, $\langle\nabla^2\tilde{c}\rangle$ vanishes by the Leibniz rule, and we have:
\begin{equation}
    \langle \partial_z w\rangle = \frac{w(z=h) - w(z=aH) }{h-aH(\vec{r}_\parallel)}.
\end{equation}
To derive an equation for $\tilde{c}$ we subtract (\ref{eq:Ceqn}) from (\ref{eq:fullCeqn}):
\begin{equation}
    \fl \partial_t \tilde{c} = - \left[\vect{\nabla} \cdot(\vec{u}C) - \left( \vect{\nabla}_\parallel \cdot(\langle \vec{u}\rangle C) +
    \langle \partial_z w\rangle C\right) \right]
    - \left[\vect{\nabla}\cdot (\vec{u}\tilde{c}) - \langle \vect{\nabla} \cdot (\vec{u}\tilde{c}) \rangle \right] + D_0 \nabla^2 \tilde{c}.
    \label{eq:ctidleeqn}
\end{equation}

As we are interested in the long-time behavior of the particle, we consider time scales greater than the time scale of diffusion in the vertical direction, $t \gsim h^2/D_0$ \cite{Stone:1999}. In this limit, diffusion across the channel averages out and we can consider the steady-state solution for $\tilde{c}$, $\partial_t \tilde{c} =0$. Further, we expect that the perturbations in the probability density are smaller than the average, $\tilde{c} \lsim C$, which leads us to neglect the second term of (\ref{eq:ctidleeqn}); this step is standard in this type of theoretical development. Finally, since $\tilde{c}$ encodes the entire fluctuations of the probability-density in the $z$-direction, we assume that the diffusion in $z$ is dominant. These assumptions lead to a simplified form of (\ref{eq:ctidleeqn}):
\begin{equation}
   D_0 \partial^2_z \tilde{c}= \vect{\nabla} \cdot(\vec{u}C) - \left( \vect{\nabla}_\parallel \cdot(\langle \vec{u}\rangle C)
   + \langle \partial_z w\rangle
   C\right).
\label{eq:ctildered}
\end{equation}

Subsequently, we rescale times by $h^2/D_0$, lengths by $h$, and velocities by $u_0$. Introducing the P{\'e}clet number $\mathrm{Pe}\equiv u_0 h/D_0$, a dimensionless roughness $\epsilon=a/h$, and exploiting incompressibility, (\ref{eq:Ceqn}) and (\ref{eq:ctildered}) become
\numparts
\begin{eqnarray}
    \partial_t C &= -\mathrm{Pe}\left[\vect{\nabla}_\parallel\cdot(\langle \vec{u}\rangle C) -\langle \partial_z w\rangle
    C - \langle \vec{u}\cdot \vect{\nabla} \tilde{c}\rangle\right] + \nabla^2_\parallel C, \label{eq:Credeqn}
    \\
    \partial^2_z\tilde{c} &= \mathrm{Pe}\left[\vec{u}\cdot\vect{\nabla}_\parallel C - \vect{\nabla}_\parallel \cdot\left(\langle \vec{u}\rangle C\right)-  \langle \partial_z w\rangle
    C \right]. \label{eq:Ctilderedreqn}
\end{eqnarray}
\endnumparts
We note here that the in-plane divergence $\nabla_\parallel\cdot\langle\vec{u}\rangle$ does not vanish in general. Specifying the surface shape $H(\vec{r}_\parallel)$ and the associated hydrodynamic flows $\vec{u}$, allows us to recast (\ref{eq:Credeqn}) into a classical, two-dimensional convection-diffusion equation with effective transport parameters.

\subsection{Perturbation flow due to surface roughness}
The base shear flow $\vec{u}_0=u_0 z/h \vec{e}_x$ that arises due to the motion of the upper boundary becomes perturbed by the presence of the surface topography. At low Reynolds numbers, the flow is governed by the continuity and Stokes equations. Non-dimensionalizing velocities $\vec{u}$ by $u_0$ and the pressure field $p$ by $\mu u_0/h$, where $\mu$ denotes the viscosity of the fluid, the governing equations are
\numparts
\begin{eqnarray}
    \nabla\cdot \vec{u} = 0, \label{eq:continuity}\\
    -\nabla p + \nabla^2 \vec{u} = \vec{0}. \label{eq:stokes}
\end{eqnarray}
\endnumparts

As we consider the case of a moving, smooth top boundary $S_h$ and a static rough bottom boundary $S_w$, the boundary conditions are $\vec{u} = \vec{u}_0 = \vec{e}_x$ on $S_h$ and $\vec{u}={\vec 0}$ on $S_w$. For small surface roughness $\epsilon=a/h\ll1$, we  can expand the flow field in terms of $\epsilon$ as
\begin{equation}
    \vec{u} = \vec{u}^{(0)} + \epsilon \vec{u}^{(1)} + \epsilon^2 \vec{u}^{(2)} + \mathcal{O}(\epsilon^3),
\end{equation}
and similarly the pressure field $ p = p^{(0)} + \epsilon p^{(1)} + \epsilon^2 p^{(2)} + \mathcal{O}(\epsilon^3)$. We include terms up to $\epsilon^2$, as we will show that these terms determine the leading-order effects of the surface roughness on dispersion. Next, we apply a domain perturbation to the boundary conditions~\cite{Kamrin:2010}, which allows us to replace the structured wall $S_w$ by a planar reference surface $S_0$, with the  boundary conditions:
\numparts
\begin{eqnarray}
\vec{u}^{(0)}&=\vec{0},\label{eq:bc_0}\\
\vec{u}^{(1)}&=-H(\vec{r}_\parallel)\partial_z\vec{u}^{(0)}|_{z=0}, \label{eq:bc_1}\\
\vec{u}^{(2)}& =-H(\vec{r}_\parallel)\partial_z\vec{u}^{(1)}|_{z=0}-\frac{H(\vec{r}_\parallel)^2}{2}\partial^2_z\vec{u}^{(0)}|_{z=0}. \label{eq:bc_2}
\end{eqnarray}
\endnumparts

From the boundary conditions we immediately find that the zeroth-order flow is simple shear $\vec{u}^{(0)} = z\vec{e}_x$ and therefore the boundary conditions reduce to:  $\vec{u}^{(1)}=-H(\vec{r}_\parallel)\vec{e}_x$ and $ \vec{u}^{(2)}=-H(\vec{r}_\parallel)\partial_z\vec{u}^{(1)}|_{z=0}$ on $S_0$.
Hence, we can model the perturbation velocity due to the presence of surface topography as an expansion in terms of $\epsilon$, with an effective slip velocity on a flat surface $S_0$ entering into the boundary conditions for every order. This slip velocity effectively encodes the presence of the surface roughness while allowing us to consider flow between two flat parallel plates. The roughness-induced velocity field can be calculated analytically for surface shapes of the form
\begin{equation}
H(\vec{r}_\parallel) = \sum_{i=1}^N \alpha_{i} \cos(\vec{k}_{i}^\alpha\cdot\vec{r}_\parallel)+ \beta_{i}\sin(\vec{k}_{i}^\beta\cdot\vec{r}_\parallel), \label{eq:3d_surface}
\end{equation}
with $N$ modes, arbitrary coefficients $\alpha_{i}$, $\beta_{i}$, and wave vectors $\vec{k}_{i}^\alpha$, $\vec{k}_{i}^\beta$ (see~\ref{appendix:flow}). This result implies that the flow fields can be obtained analytically for any shape function expandable in terms of a Fourier series.

\begin{figure}[htp]
\centering
    \includegraphics[width=\linewidth]{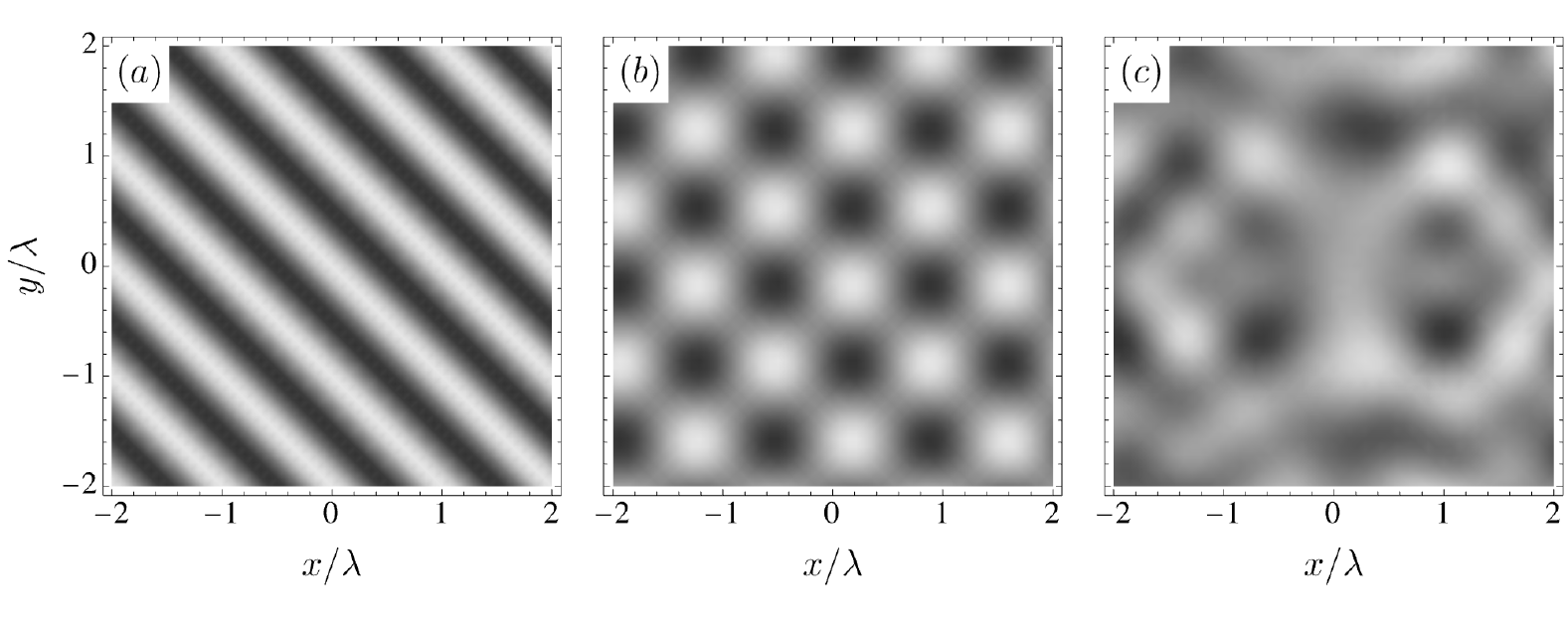}
    \caption{Different surface structures. (a) Corrugated surface  $H(\vec{r}_{\parallel})=\cos(\vec{k}\cdot \vec{r}_\parallel)$ with $\vec{k}=(2\pi/\lambda) [\cos(\pi/4),\sin(\pi/4),0]^{\rm T}$, (b) bumpy surface $H(\vec{r}_\parallel)=(\cos(\vec{k}_0\cdot\vec{r}_\parallel)+\cos(\vec{k}_1\cdot\vec{r}_\parallel))/2$ with $\vec{k}_0=(2\pi/\lambda) [\cos(\pi/4),\sin(\pi/4),0]^{\rm T}$ and $\vec{k}_1=(2\pi/\lambda) [\cos(\pi/4),-\sin(\pi/4),0]^{\rm T}$, and (c) a realization of a randomly structured surface as described in (\ref{eq:random_surface}) with $N=20$ modes. Black indicates a surface height of $1$ and white of $-1$. }
    \label{fig:surfaces}
\end{figure}

We next consider a limiting case of (\ref{eq:3d_surface}) of a simple corrugated surface, which are commonly used in microfluidic settings~\cite{Stroock:2002} to help us develop the form of the dispersion tensor. Following this, to study tracer dispersion, we also consider more complex forms of (\ref{eq:3d_surface}), including periodic bumpy as well as randomly structured surfaces. See \fref{fig:surfaces} for exemplary surface structures.

\subsection{Corrugated Surface (single mode)}
\subsubsection{Details of the flow field} We start our analysis by considering a surface shape consisting of a single Fourier mode
\begin{equation}
    H(\vec{r}_\parallel) = \cos(\vec{k}\cdot\vec{r}_\parallel),
\end{equation}
where $\vec{k}=[k_x,k_y,0]^{\rm T}$ denotes the in-plane wave vector [\fref{fig:surfaces}(a)]. Due to the boundary condition $\vec{u}^{(1)}|_{z=0}=-\cos(\vec{k}\cdot\vec{r}_\parallel)\vec{e}_x$ (see (\ref{eq:bc_1})) the first-order flow fields must take the form
\numparts
\begin{eqnarray}
    u^{(1)}(\vec{r}; \vec{k}) &=\hat{u}^{(1)}(z; \vec{k})\cos(\vec{k}\cdot\vec{r}_\parallel),\\
    v^{(1)}(\vec{r}; \vec{k})  &=\hat{v}^{(1)}(z; \vec{k})\cos(\vec{k}\cdot\vec{r}_\parallel),\\
    w^{(1)}(\vec{r}; \vec{k})  &= \hat{w}^{(1)}(z; \vec{k})\sin(\vec{k}\cdot\vec{r}_\parallel),
\end{eqnarray}
\endnumparts
and consequently the pressure can be written as $p^{(1)}(\vec{r}; \vec{k}) = \hat{p}^{(1)}(z; \vec{k})\sin(\vec{k}\cdot\vec{r}_\parallel)$. Inserting this ansatz into the continuity equation (\ref{eq:continuity}), we can write the relationship between the $z-$component of the velocity field and the $x-$ and $y-$components via
\begin{eqnarray}
    \frac{\rmd \hat{w}^{(1)}}{\rmd z} &= k_x\hat{u}^{(1)}+k_y\hat{v}^{(1)} = \vec{k}\cdot[\hat{u}^{(1)}, \hat{v}^{(1)},0]^{\rm T}. \label{eq:w1}
\end{eqnarray}
Using this relation and the Stokes equations (\ref{eq:stokes}) we can obtain analytical expressions for the velocity field using Mathematica (see~\ref{appendix:flow}). It is convenient to introduce a vector field $\vec{Q}(z; \vec{k})= [Q_1,Q_2,0]^{\rm T}$, which is related to the velocity field via $\rmd^2 Q_1/\rmd z^2= \hat{u}^{(1)}$ and $\rmd^2 Q_2/\rmd z^2= \hat{v}^{(1)}$. Then, using (\ref{eq:w1}), the first-order roughness-induced velocities can be expressed in terms of $\vec{Q}$ via
\begin{eqnarray}
    \vec{u}^{(1)}(\vec{r}; \vec{k})&= \cos(\vec{k}\cdot\vec{r}_\parallel)\vec{Q}_{zz}+\sin(\vec{k}\cdot\vec{r}_\parallel)(\vec{k}\cdot\vec{Q}_z)\vec{e}_z, \label{eq:vel1_cos}
\end{eqnarray}
where the subscript $z$ denotes the derivative with respect to $z$. We note that the first-order flow fields are periodic functions of the in-plane coordinates $x$ and $y$.

From the first-order velocities, the second-order boundary condition on $z=0$ (\ref{eq:bc_2}) becomes
\begin{eqnarray}
    \vec{u}^{(2)}(\vec{r}; \vec{k})\bigl|_{z=0} &= -\frac{1}{2}(1+\cos(2\vec{k}\cdot\vec{r}_\parallel))\vec{Q}_{zzz}\bigl|_{z=0}
    +\frac{1}{2}\sin(2\vec{k}\cdot\vec{r}_\parallel)(\vec{k}\cdot\vec{e}_x)\vec{e}_z,\nonumber
\\
    &\equiv \bar{\vec{u}}_0^{(2)}(\vec{k})+\tilde{\vec{u}}^{(2)}(\vec{r}; \vec{k})|_{z=0},\label{eq:bc2_np}
\end{eqnarray}
where we have written the second-order boundary condition as a combination of a non-periodic, $\bar{\vec{u}}^{(2)}_0(\vec{k})$, and a periodic, $\tilde{\vec{u}}^{(2)}(\vec{r}; \vec{k})$, contribution, which is periodic over the spatial coordinate $\vec{r}_\parallel$. Therefore, the second-order flow field $\vec{u}^{(2)}(\vec{r}; \vec{k})$ assumes the general, abbreviated form
\begin{eqnarray}
    \vec{u}^{(2)}(\vec{r}; \vec{k}) = \bar{\vec{u}}_0^{(2)}(\vec{k})(1-z)+\tilde{\vec{u}}^{(2)}(\vec{r};\vec{k}). \label{eq:vel2_cos}
\end{eqnarray}
We note that the $z-$component of the velocity field is completely periodic with zero mean and hence $\bar{w}_0^{(2)}(\vec{k})=0$. The expressions for the velocity fields have been obtained analytically using Mathematica. The non-periodic contributions, $\bar{\vec{u}}^{(2)}_0(\vec{k})$, can be found in~\ref{appendix:flow}.

\subsubsection{Derivation of the drift-diffusion equation}
In what follows we present a detailed derivation of the drift-diffusion equation for a 3D corrugated surface shape. We start by substituting the velocity field in (\ref{eq:vel1_cos}) and (\ref{eq:vel2_cos}) into (\ref{eq:Credeqn})-(\ref{eq:Ctilderedreqn}):
\numparts
\begin{eqnarray} \eqalign{\fl \partial_t C = \mathrm{Pe}\left[-\frac{1}{2}\left(\partial_x + \epsilon^2\bar{\vec{u}}_0^{(2)}\cdot \vect{\nabla}_\parallel \right)C - \left\langle \left(z \partial_x + (\epsilon \vec{u}^{(1)} + \epsilon^2\vec{u}^{(2)})\cdot \vect{\nabla} \right)\tilde{c}\right\rangle \right] +\nabla^2_\parallel C\\+ \mathrm{p.t.} +\mathcal{O}(\epsilon^3), \label{eq:Cwave} } \end{eqnarray}
\begin{eqnarray}
   \eqalign{\fl
   \partial_z^2 \tilde{c} = \mathrm{Pe}\Bigg[\left(z-\frac{1}{2}\right)\partial_x  + \epsilon \left(\vec{u}^{(1)} \cdot \vect{\nabla}_\parallel - \vect{\nabla}_\parallel\cdot \left\langle \vec{u}^{(1)}\right\rangle\right)
   + \epsilon^2\left(\left(\frac{1}{2}-z\right) \bar{\vec{u}}^{(2)}_0 \cdot \vect{\nabla}_\parallel \right)\Bigg]C \\ + \mathrm{p.t.} +\mathcal{O}(\epsilon^3).}
   \label{eq:Ctildewave}
\end{eqnarray}
\endnumparts
Here, `p.t.' (periodic terms) includes all terms that vary periodically over the horizontal coordinate $\vec{r}_\parallel$. As we are concerned primarily with the long-time, average diffusivity of the particles we only consider the non-periodic contributions to the particle concentration. All of the periodic effects will average out when considering macroscopic motion, and so we collect all of these terms in `p.t.' for further calculations. At this stage, this includes several terms of $\mathcal{O}(\epsilon^2)$ as well as the term $\epsilon \vect{\nabla}_\parallel\cdot(\langle \vec{u}^{(1)} \rangle C)$ in (\ref{eq:Cwave}). While some periodic terms included in p.t. in $\tilde{c}$ may lead to non-periodic contributions to $C$, such terms will be $\mathcal{O}(\epsilon^3)$ or smaller and so can be neglected.

We proceed by integrating (\ref{eq:Ctildewave}) twice in $z$ to obtain
\begin{eqnarray}\label{eq:ctilde_int}
\fl
    \tilde{c} = \mathrm{Pe}\Bigg[\left(\frac{z^3}{6}-\frac{z^2}{4}\right)\partial_x + \epsilon \Bigg(\cos(\vec{k}\cdot\vec{r}_\parallel) \left(\vec{Q}\cdot\vect{\nabla}_\parallel  - \frac{z^2}{2}\langle \vec{Q}_{zz}\rangle\cdot\vect{\nabla}_\parallel\right) + \sin(\vec{k}\cdot\vec{r}_\parallel)\frac{z^2}{2}\langle \vec{Q}_{zz}\rangle\cdot\vec{k} \Bigg)\nonumber \\ - \epsilon^2 \left(\frac{z^3}{6}-\frac{z^2}{4}\right) \bar{\vec{u}}_0^{(2)} \cdot \vect{\nabla}_\parallel \Bigg]C + \lambda_1 z + \lambda_2 + \mathrm{p.t.} + \mathcal{O}(\epsilon^3)
\end{eqnarray}
where $\lambda_1$ and $\lambda_2$ are constants of integration. By taking the height average of the continuity equation, we find that
\begin{eqnarray} \label{eq:continuity0}
 \left\langle\vec{Q}_{zz}\right\rangle\cdot\vec{k} = \left\langle\partial_z \tilde{w}^{(1)}(z) \right\rangle= \tilde{w}^{(1)}(1)-\tilde{w}^{(1)}(0)=0,
\end{eqnarray}
which simplifies (\ref{eq:ctilde_int}).
The no-flux boundary conditions, $\partial_z \tilde{c} = 0$ on $z = 0$ and $z=1$, provide the same condition for $\lambda_1$:
\begin{equation}
    \lambda_1 = -\epsilon \cos(\vec{k}\cdot\vec{r}_\parallel) \vec{Q}_z(z=0) \cdot \vect{\nabla}_\parallel C + \mathrm{p.t.}+ \mathcal{O}(\epsilon^3). \label{eq:lambda1}
\end{equation}
As by definition the vertical average vanishes, i.e., $\langle \tilde{c}\rangle = 0$,  $\lambda_2$ is
\begin{eqnarray}
\eqalign{\fl
    \lambda_2  = \mathrm{Pe}\Bigg[ \frac{1}{24}\partial_x - \epsilon \cos(\vec{k}\cdot\vec{r}_\parallel) \bigg(\langle \vec{Q}\rangle- \frac{1}{6}\langle \vec{Q}_{zz}\rangle
    -\frac{1}{2}\vec{Q}_z(0)\bigg)\cdot\vect{\nabla}_\parallel  - \frac{\epsilon^2}{24} \bar{\vec{u}}_0^{(2)} \cdot \nabla_\parallel\Bigg]C \\ + \mathrm{p.t.} + \mathcal{O}(\epsilon^3).}
\end{eqnarray}
Thus, the perturbed probability density $\tilde{c}(\vec{r},t; \vec{k})$ is given by
\begin{equation} \label{eq:ctilde_final}
\fl
\tilde{c} = \mathrm{Pe}\big[g(z) \partial_x + \epsilon \cos(\vec{k}\cdot\vec{r}_\parallel)\vec{f}\cdot \vect{\nabla}_\parallel
- \epsilon^2 g(z) \bar{\vec{u}}_0^{(2)} \cdot \vect{\nabla}_\parallel \big]C + \mathrm{p.t.} + \mathcal{O}(\epsilon^3),
\end{equation}
where we abbreviated the vector field
\begin{equation}
\fl
    \vec{f}(z; \vec{k}) = \vec{Q}(z; \vec{k}) - \left(\frac{z^2}{2}-\frac{1}{6}\right)\langle \vec{Q}_{zz}(z;\vec{k})\rangle
    -\langle \vec{Q}(z;\vec{k}) \rangle - \left(z-\frac{1}{2}\right)\vec{Q}_z(z=0;\vec{k}).
\end{equation}
and introduced the function
\begin{equation}
    g(z) = \frac{z^3}{6}-\frac{z^2}{4} + \frac{1}{24}.
\end{equation}

To derive the effective drift-diffusion equation up to second order in the surface roughness $\mathcal{O}(\epsilon^2)$, we insert (\ref{eq:ctilde_final}) into (\ref{eq:Credeqn}) and compute the contributions of the terms $\left\langle \left(z \partial_x + (\epsilon \vec{u}^{(1)} + \epsilon^2\vec{u}^{(2)})\cdot \vect{\nabla}\right)\tilde{c}\right\rangle$ to $\mathcal{O}(\epsilon^2)$. The $\mathcal{O}(\epsilon^2)$ contribution to the first term is
\begin{equation}
    \langle z \partial_x \tilde{c}\rangle = \frac{\mathrm{Pe}}{120}\left[-\partial_x^2 + \epsilon^2\bar{\vec{u}}_0^{(2)} \cdot \vect{\nabla}_\parallel\partial_x \right]C + \mathrm{p.t.} + \mathcal{O}(\epsilon^3).
    \label{eq:order2}
\end{equation}
We further calculate non-vanishing terms to $\mathcal{O}(\epsilon)$ of
\begin{equation}
     \langle \vec{u}^{(1)} \cdot \vect{\nabla} \tilde{c}\rangle = \frac{1}{2}\epsilon\mathrm{Pe}\left(\vect{\nabla}_\parallel \cdot \left\langle \vec{Q}_{zz} \vec{f}\right\rangle\cdot\vect{\nabla}_\parallel\right) C + \mathrm{p.t.} + \mathcal{O}(\epsilon^2),
\end{equation}
and terms to $\mathcal{O}(1)$ of
\begin{equation}
    \langle \vec{u}^{(2)} \cdot \vect{\nabla} \tilde{c} \rangle = \frac{\mathrm{Pe}}{120}\left(\bar{\vec{u}}^{(2)}_0 \cdot \vect{\nabla}_\parallel\right) \partial_x C + \mathrm{p.t.} + \mathcal{O}(\epsilon).
    \label{eq:order0}
\end{equation}
By inserting these terms into (\ref{eq:Cwave}), the  convection-diffusion (Fokker-Planck) equation (\ref{eq:fokkerplank}) reduces to a 2D problem when considering only the non-periodic effects. Therefore, we will drop the $z$ component of the vectors $\vect{\nabla}_\parallel$, $\vec{r}_\parallel$, and $\vec{k}$, henceforth treating them as two-dimensional quantities.

Similar to previous work on dispersion through spatially periodic, porous media~\cite{Brenner:1980}, we can divide the domain into subdomains of length $L_x$ along the $x-$direction and $L_y$ along the $y-$direction, where the length scales $L_x$ and $L_y$ are set by the underlying periodic surface structure. The drift-diffusion equation contains periodic terms (p.t.), which vary locally within one subdomain as a result of the periodic flows generated by the surface structure, and non-periodic terms, which are constant in the spatial coordinate $\vec{r}_\parallel$. We will consider (here in dimensional form) coarse-grained length scales $x,y\gg L_x,L_y$ and long times $t\gg {\rm max} (L_x,L_y)/(\epsilon u_0)$, so that the local effects of the periodic terms can be neglected, thus dropping all `p.t.' from further consideration.

The 2D convection-diffusion equation for the concentration (probability density) $C\equiv C(\vec{r}_\parallel,t;\vec{k})$ of tracer particles considered on these course-grained scales is
\begin{equation}
    \partial_t C = - \vect{\nabla}_\parallel \cdot\left(\vec{u}_{\rm eff} C\right) + \vect{\nabla}_\parallel \cdot \left(\bm{\mathrm{D}}\cdot \vect{\nabla}_\parallel C\right) +\mathcal{O}(\epsilon^3), \label{eq:asymptotic_FB}
\end{equation}
with effective drift velocity and dispersion tensor:
\numparts
\begin{eqnarray}
\vec{u}_{\rm eff} &= \vec{u}_{\rm eff}^{(0)} + \epsilon^2 \vec{u}_{\rm eff}^{(2)} +\mathcal{O}(\epsilon^3), \label{eq:u_eff}\\
\vec{D} &= \vec{D}^{(0)} + \epsilon^2 \vec{D}^{(2)}+\mathcal{O}(\epsilon^3).\label{eq:dispersion}
\end{eqnarray}
\endnumparts
Note that we have defined $\vec{u}_{\rm eff}$ and $\vec{D}$ to include the dependence on P{\'e}clet number. In agreement with earlier work, the zeroth-order contributions to the effective drift velocity and dispersion are:
 \numparts
 \begin{equation}
 \vec{u}_{\rm eff}^{(0)}=\frac{1}{2}{\rm Pe}\left[\begin{array}{c}1 \\0\end{array}\right] \label{eq:u0}
 \end{equation}
 \begin{equation}
     \vec{D}^{(0)} =
	\left[
	\begin{array}{cc}
		 1 + \frac{{\rm \rm Pe}^2}{120} & 0 \\
		0 & 1
	\end{array}
	\right]. \label{eq:diff0}
\end{equation}
 \endnumparts
The roughness-induced contributions, which represent a main finding of this work, evaluate to:
\begin{eqnarray}
 \vec{u}_{\rm eff}^{(2)}(\vec{k})=\frac{1}{2}{\rm Pe}\left[\begin{array}{c}\bar{u}_0^{(2)}(\vec{k})\\
\bar{v}_0^{(2)}(\vec{k})\end{array}\right],\label{eq:u2}
\end{eqnarray}
 and
\begin{eqnarray}
    \label{eq:deff_2}
    \setlength{\arraycolsep}{5pt}
    \renewcommand{\arraystretch}{1.5}
	\eqalign{\fl \vec{D}^{(2)}(\vec{k}) = - \frac{{\rm Pe}^2}{4} \left(\frac{1}{15}
	\left[
	\begin{array}{cc}
		\bar{u}_0^{(2)}(\vec{k})  & \frac{1}{2}\bar{v}_0^{(2)}(\vec{k}) \\
		\frac{1}{2}\bar{v}_0^{(2)}(\vec{k})  & 0
	\end{array}
	\right] \right. \\ \left.
        \vphantom{\begin{array}{cc}
		\bar{u}_0^{(2)}(\vec{k})  \\
		\frac{1}{2}\bar{v}_0^{(2)}(\vec{k})
	\end{array}}
        +\left(\left\langle \vec{Q}_{zz}(z;\vec{k}) \vec{f}(z;\vec{k}) \right\rangle +\left\langle \vec{Q}_{zz}(z;\vec{k})  \vec{f}(z;\vec{k}) \right\rangle^{\rm T}\right)
	\right).}
\end{eqnarray}

 \begin{figure}[bt]
    \centering
    \includegraphics[width=0.6\linewidth]{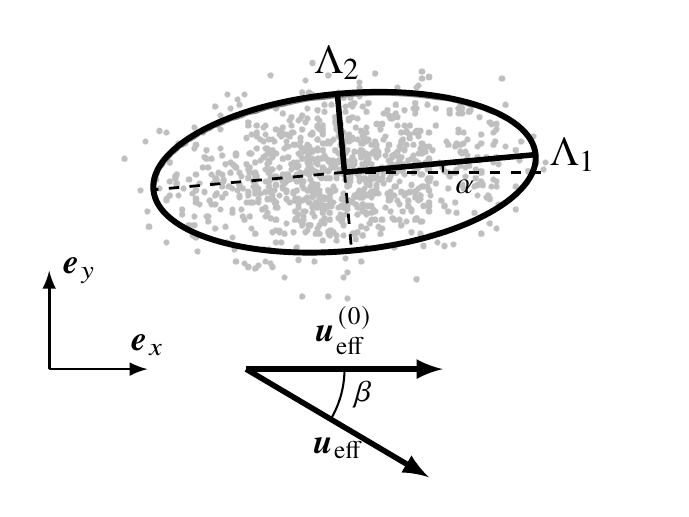}
    \caption{Sketch of a snapshot of the distribution of particles diffusing in shear flow near a corrugated surface with $\vec{k} = [2\pi,2\pi]^\mathrm{T}$  at long times. Here the probability distribution follows a bivariate Gaussian distribution (\ref{eq:bivariate}), where $\Lambda_1$ and $\Lambda_2$ denote the eigenvalues of the dispersion matrix, the black lines indicate the directions of the eigenvectors, and the ellipse represents a level curve of the distribution equal to $0.1 C_{\rm max}$. The eigenvector corresponding to $\Lambda_1$ is at an angle $\alpha$ with the applied forcing. Further, $\beta$ denotes the angle between the drift velocity and the direction $\vec{e}_x$, which is also the direction for the average flow in the case of a planar wall. Note that for the particle distribution and drift velocity the vertical scale is stretched by a factor of 25 for clarity.
    }
    \label{fig:ellipses}
\end{figure}

The effective drift and dispersion tensor are functions of the wave vector $\vec{k}$, the roughness $\epsilon$, and the P{\'e}clet number (Pe). Note that only even powers in $\epsilon$ contribute to the effective drift and dispersion tensors. In particular, the zeroth-order contributions correspond to the classical Taylor dispersion in shear flow between two planar walls and the second-order contribution reflects the effects of the surface topography.

The form of the effective drift suggests that surface structures can induce particles to drift in directions that are different from the direction of applied forcing (flow), as surface corrugations can deflect the average flow direction by an angle $\beta$, see \fref{fig:ellipses}.

Furthermore, we  observe that the roughness-induced dispersion tensor is symmetric and contains non-vanishing off-diagonal terms, which represent an $xy-$coupling. This indicates that the hydrodynamic interactions of the fluid flow with the surface structure not only alters the dispersion of tracer particles but also changes the nature of the long-time anisotropic diffusion. In particular, we note that the solution $C$ to the coarse-grained convection-diffusion equation (\ref{eq:asymptotic_FB}) assumes a bivariate Gaussian distribution
\begin{eqnarray}
C = \frac{1}{\mathcal{N}}\exp\left[-\frac{1}{2} \left(\vec{r}_\parallel-\boldsymbol{\mu}\right)^{\rm T}\boldsymbol{\cdot}\boldsymbol{\Sigma}^{-1}\boldsymbol{\cdot}\left(\vec{r}_\parallel-\boldsymbol{\mu}\right)\right], \label{eq:bivariate}
\end{eqnarray}
with normalization $\mathcal{N}=2\pi \sqrt{{\rm det}(\boldsymbol{\Sigma})}$,  mean $\boldsymbol{\mu} = \vec{u}_{\rm eff} t$, and variance
$\boldsymbol{\Sigma} = 2 t \vec{D}$ with $\vec{u}_{\rm eff}$ and $\vec{D}$ from (\ref{eq:u_eff})-(\ref{eq:dispersion}). The level sets of the distribution are ellipsoids with major and minor axis along the eigenvectors of the dispersion matrix, which are proportional to the respective eigenvalue, $\Lambda_1$ and $\Lambda_2$, see \fref{fig:ellipses}. Hence, the particle diffuses along the eigenvectors of the dispersion matrix with diffusivities, $\Lambda_1$ and $\Lambda_2$ ($\Lambda_1\geq\Lambda_2$).

It is important to note that the eigenvectors of the dispersion matrix generally do not align with the effective drift, $\vec{u}_{\rm eff}^{(0)}+\epsilon^2\vec{u}_{\rm eff}^{(2)}$, as one may naively expect. The angle between the principle eigenvector and the drift, $\alpha + \beta$, where $\alpha$ is the angle between the applied forcing and the eigenvector corresponding to~$\Lambda_1$ and $\beta$ the deflection of the average flow from the forcing (see \fref{fig:ellipses}) depend on details of the surface structure.

We discuss results for a single surface mode in Sec.~\ref{sec:results_1mode}.

\subsection{Bumpy Surface (two modes)}
We can further generalize this approach by considering a bumpy surface
\begin{equation}
    H(\vec{r}_\parallel) = H_1 \cos(\vec{k}_{1}\cdot \vec{r}_\parallel) + H_2 \sin(\vec{k}_{2}\cdot \vec{r}_\parallel),
    \label{eq:heightsbumps}
\end{equation}
where $\vec{k}_1$, $\vec{k}_2$ are parallel to the surface $S_0$ and $H_1$ and $H_2$ are two dimensionless parameters.
Following our previous notation, the horizontal components of the first-order velocity $\vec{u}^{(1)}$ assume the form
\begin{eqnarray}
\fl
    \eqalign{\vec{u}^{(1)}(\vec{r};\{H_i,\vec{k}_i\}) = &H_1\cos(\vec{k}_1\cdot\vec{r}_\parallel)\vec{Q}_{zz}(z; \vec{k}_1)
    +H_2\sin(\vec{k}_{2}\cdot \vec{r}_\parallel)\vec{Q}_{zz}(z; \vec{k}_2)\\
    &+\left( H_1\sin(\vec{k}_1\cdot\vec{r}_\parallel)\vec{k}_1\cdot \vec{Q}_z(z; \vec{k}_1)-H_2\cos(\vec{k}_{2}\cdot \vec{r}_\parallel)\vec{k}_2\cdot\vec{Q}_z(z;\vec{k}_2)\right)\vec{e}_z.}
\end{eqnarray}

Consequently, the second-order boundary conditions on $z=0$ reduce to
\begin{equation}
    \vec{u}^{(2)}(\vec{r}; \{H_i,\vec{k}_i\})\bigl|_{z=0}  = -\frac{1}{2}\sum_{i=1}^2(H_i)^2 \vec{Q}_{zzz}(z=0; \vec{k}_i) + {\rm p.t.}
\end{equation}
where `p.t.' contains all periodic terms. Note that the non-periodic terms do not include a coupling term $H_1H_2$. The second-order velocities then assume the form
\begin{equation}
     \vec{u}^{(2)}(\vec{r}; \{H_i,\vec{k}_i\})
     = \bar{\vec{u}}_0^{(2)}(\{H_i,\vec{k}_i\})(1-z)
     +\tilde{\vec{u}}^{(2)}(\vec{r};\{H_i,\vec{k}_i\}), \label{eq:vel2_bumps}
\end{equation}
with $\bar{\vec{u}}_0^{(2)}(\{H_i,\vec{k}_i\})=\sum_{i=1}^2H_i^2\bar{\vec{u}}_0^{(2)}(\vec{k}_i)$.
The derivation of the diffusion tensor proceeds as in the case of a single mode surface up to (\ref{eq:Cwave}) and (\ref{eq:Ctildewave}). Due to the presence of the additional terms in the second-order flow, the equation for $\tilde{c}$ becomes
\begin{eqnarray}
\eqalign{
\fl
\tilde{c} = \mathrm{Pe}\Bigg[g(z)\partial_x + \epsilon \big(H_1 \cos(\vec{k}_1\cdot \vec{r}_\parallel)\vec{f}(z;\vec{k}_1) + H_2 \sin(\vec{k}_2\cdot \vec{r}_\parallel) \vec{f}(z;\vec{k}_2)\big)\cdot \vect{\nabla}_\parallel\\ - \epsilon^2 g(z) \left(\sum_{i=1}^2 H_i^2 \bar{\vec{u}}_0^{(2)}(\vec{k}_i)\right) \cdot \vect{\nabla}_\parallel \Bigg]C + \mathrm{p.t.} + \mathcal{O}(\epsilon^3),}
\end{eqnarray}
where the contributions of different modes to the velocity fields have been summed (compare with (\ref{eq:ctilde_int})) and the term $\langle \vec{Q}_{zz}(z;\vec{k}_{1,2})\rangle\cdot\vec{k}_{1,2}=0$ vanishes for each mode. We further replace $\bar{\vec{u}}_0^{(2)}$ in (\ref{eq:order2}) and (\ref{eq:order0}) by $\bar{\vec{u}}_0^{(2)}(\{H_i,\vec{k}_i\})$ (from (\ref{eq:vel2_bumps})) and obtain the non-vanishing terms up to $\mathcal{O}(\epsilon)$ of
\begin{equation}
\fl
     \langle \vec{u}^{(1)} \cdot \vect{\nabla} \tilde{c}\rangle =
     \frac{1}{2}\epsilon\mathrm{Pe}\Bigg(\sum_{i=1}^2 H_i^2\vect{\nabla}_\parallel \cdot \langle \vec{Q}_{zz}(z;\vec{k}_i) \vec{f}(z;\vec{k}_i)\rangle  \cdot\vect{\nabla}_\parallel\Bigg) C + \mathrm{p.t.} + \mathcal{O}(\epsilon^3),
\end{equation}
Thus, the equation for the concentration $C\equiv C(\vec{r}_\parallel, t;\{H_i,\vec{k}_i\})$ in 2D reduces to (\ref{eq:asymptotic_FB}),
 where the zeroth-order contributions are the same as before (\ref{eq:u0})-(\ref{eq:diff0}) and the roughness-induced drift velocity and dispersion tensor are:
 \numparts
 \begin{equation}
 \setlength{\arraycolsep}{5pt}
    \renewcommand{\arraystretch}{1.5}
 \vec{u}_{\rm eff}^{(2)}=\frac{1}{2}{\rm Pe}\sum_{i=1}^2H_i^2 \left[\begin{array}{c} \bar{u}_0^{(2)}(\vec{k}_i)\\
 \bar{v}_0^{(2)}(\vec{k}_i)
 \end{array}\right]
 = \sum_{i=1}^2 H_i^2\vec{u}_{\rm eff}^{(2)}(\vec{k}_i)
 \end{equation}
 \begin{equation}
    \vec{D}^{(2)} = \sum_{i=1}^2 H_i^2\vec{D}^{(2)}(\vec{k}_i).
    \label{eq:diffBumps}
 \end{equation}
\endnumparts
We find that the second-order contributions are the sums over the effective drift (\ref{eq:u0}) and dispersion (\ref{eq:deff_2}) caused by a single wavy perturbation with wave vector $\vec{k}_i$. Coupling of surface modes, $H_1H_2$, is only relevant when examining effects at higher orders.

\subsection{General, randomly structured surface}
More generally, we consider the surface shape of the form of a discrete Fourier series
\begin{equation}
H(\vec{r}_\parallel) = \frac{1}{N}\sum_{i,j=1}^N \alpha_{ij} \cos(\vec{k}_{ij}\cdot\vec{r}_\parallel)+ \beta_{ij}\sin(\vec{k}_{ij}\cdot\vec{r}_\parallel), \label{eq:random_surface}
\end{equation}
with $\vec{k}_{ij}=2\pi [i,j,0]^{\rm T}/(\lambda N)$, where $\lambda$ denotes the characteristic (smallest) wavelength of the surface and $\lambda_N=\lambda N$ the longest wavelength. The latter sets the periodicity of the surface along the $x-$ and $y-$directions, respectively, and hence the coarse scales $L_x =L_y=\lambda_N$. The coefficients $\{\alpha_{ij}, \beta_{ij}\}$ are non-dimensional constants.
We can extend our procedure for the case of two modes to the general case of a surface with arbitrarily many modes, and obtain the analytical predictions for the roughness-induced effective velocities and dispersion:
\numparts
\begin{eqnarray}
 \vec{u}_{\rm eff}^{(2)}  = \frac{1}{N^2}\sum_{i,j=1}^N\left((\alpha_{ij})^2 + (\beta_{ij})^2\right)\vec{u}_{\rm eff}^{(2)}(\vec{k}_{ij}),\\
\bm{\mathrm{D}}^{(2)}  = \frac{1}{N^2} \sum_{i,j=1}^N\left((\alpha_{ij})^2+(\beta_{ij})^2\right)\vec{D}^{(2)}(\vec{k}_{ij}). \label{eq:randsurfD}
\end{eqnarray}
\endnumparts
Our findings demonstrate that the roughness-induced contributions to the dispersion at second order can be separated into linearly independent components for each surface mode.

\section{Results and Discussion}

We now use the theoretical structure developed in the previous section to examine the effects of surface topography on particle transport in a few particular cases of interest, including a corrugated surface consisting of a single mode, a bumpy surface composed of two balanced modes, and a randomly structured surface.

\subsection{Corrugated surface (1 mode) \label{sec:results_1mode}}
First, we consider surfaces that have a corrugation wavelength $\lambda$ and are oriented relative to the applied forcing direction $\vec{e}_x$ by an angle $\theta$. This corresponds to a mode with wave vector $\vec{k} = \frac{2\pi}{\lambda}[\cos\theta,\sin\theta]^{\rm T}$.
\fref{fig:simulations_dist} depicts results of our Brownian dynamics simulations (see~\ref{appendix:BD}) and theoretical predictions for a planar surface and a corrugated surface (with $\lambda/h = 1/\sqrt{2}$, $\theta=\pi/4$, and $\epsilon = 0.1$; see \fref{fig:ellipses}~(a)) with ${\rm Pe}~=100$. Compared to the planar surface, which represents classical Taylor dispersion, the presence of surface corrugations causes the particles to drift in the $-\vec{e}_y$ direction. Furthermore,  the axes of the elliptical tracer distribution tilt relative to the lab coordinate system, which is consistent with our prediction of an anisotropic dispersion tensor with non-vanishing off-diagonal elements. The level curves of the coarse-grained analytical prediction for $C$ (\ref{eq:bivariate}) capture the simulation results well.

\begin{figure}
    \centering
    \includegraphics[width=.5\linewidth]{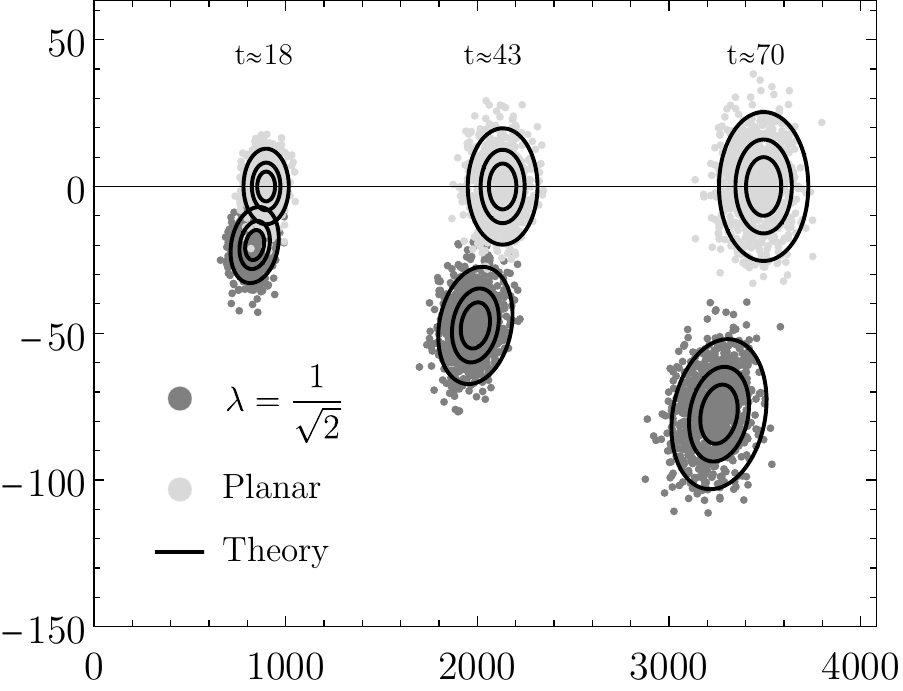}
    \caption{Distribution of tracer particles at three different times for the case of a planar (flat) wall and a corrugated wall with $\lambda = 1/\sqrt{2}$, oriented at $\theta=\pi/4$ relative to the forcing direction $\vec{e}_x$, and $\epsilon=0.1$. Here, the P{\'e}clet number is Pe~$=100$. The black lines denote the theoretical predictions for the level sets of the concentration of particles at $0.1 C_{\rm max}$, $0.4 C_{\rm max}$, and $0.7 C_{\rm max}$, where $C_{\rm max}$ is the supremum of (\ref{eq:bivariate}).
    \label{fig:simulations_dist}}
\end{figure}
\begin{figure*}[htp]
\centering
\includegraphics[width = \linewidth]{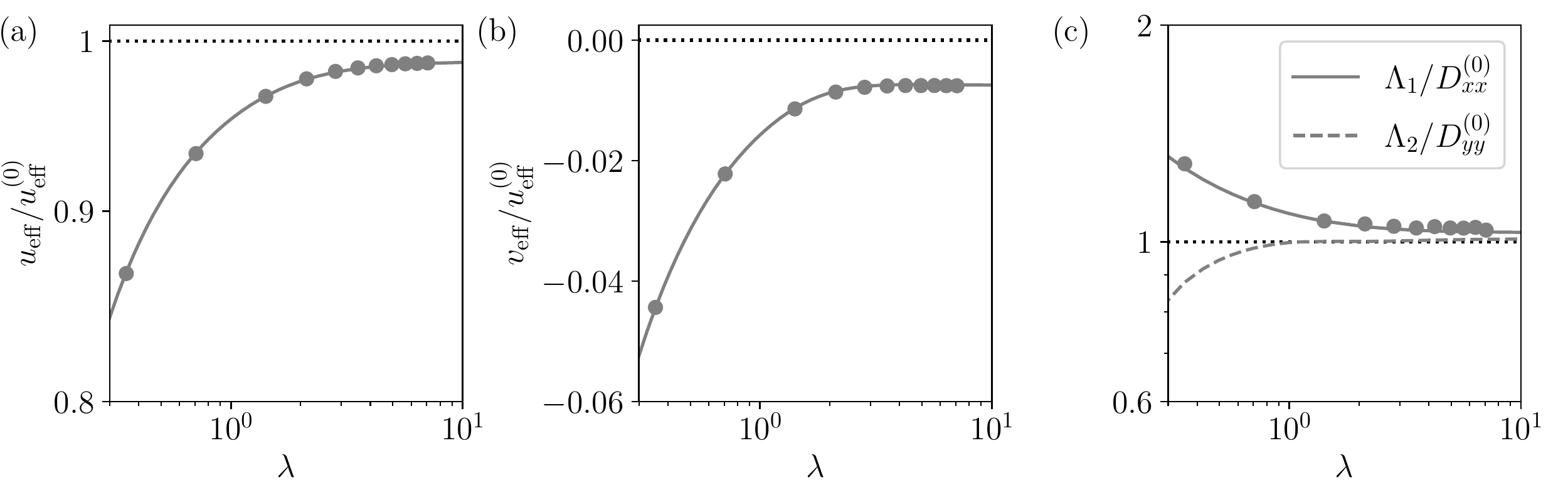}
\caption{Effective transport coefficients
as a function of wavelength~$\lambda$. (a) and (b) show the components of the effective drift velocity, $\vec{u}_{\rm eff} = [u_{\rm eff}, v_{\rm eff}]^{\rm T}$, respectively, and (c) displays the eigenvalues of the dispersion matrix, $\Lambda_1$ and $\Lambda_2$. Lines indicate theoretical predictions and symbols are results from Brownian dynamics simulations. Results are normalized by the predictions for classical Taylor dispersion, $D_{xx}^{(0)}$ and $D_{yy}^{(0)}$ (\ref{eq:diff0}), and the base flow, $u^{(0)}_{\rm eff}$ (\ref{eq:u0}). The surface is tilted at an angle $\theta=\pi/4$ and the roughness is $\epsilon=0.1$. Here, the P{\'e}clet number is Pe~$=100$. Note that we cannot adequately resolve $\Lambda_2$ in our simulations as it is of the order $\mathcal{O}(10^{-3})$.\label{fig:simulations}}
\end{figure*}
\begin{figure*}[htp]
    \centering
    \includegraphics[width=\linewidth]{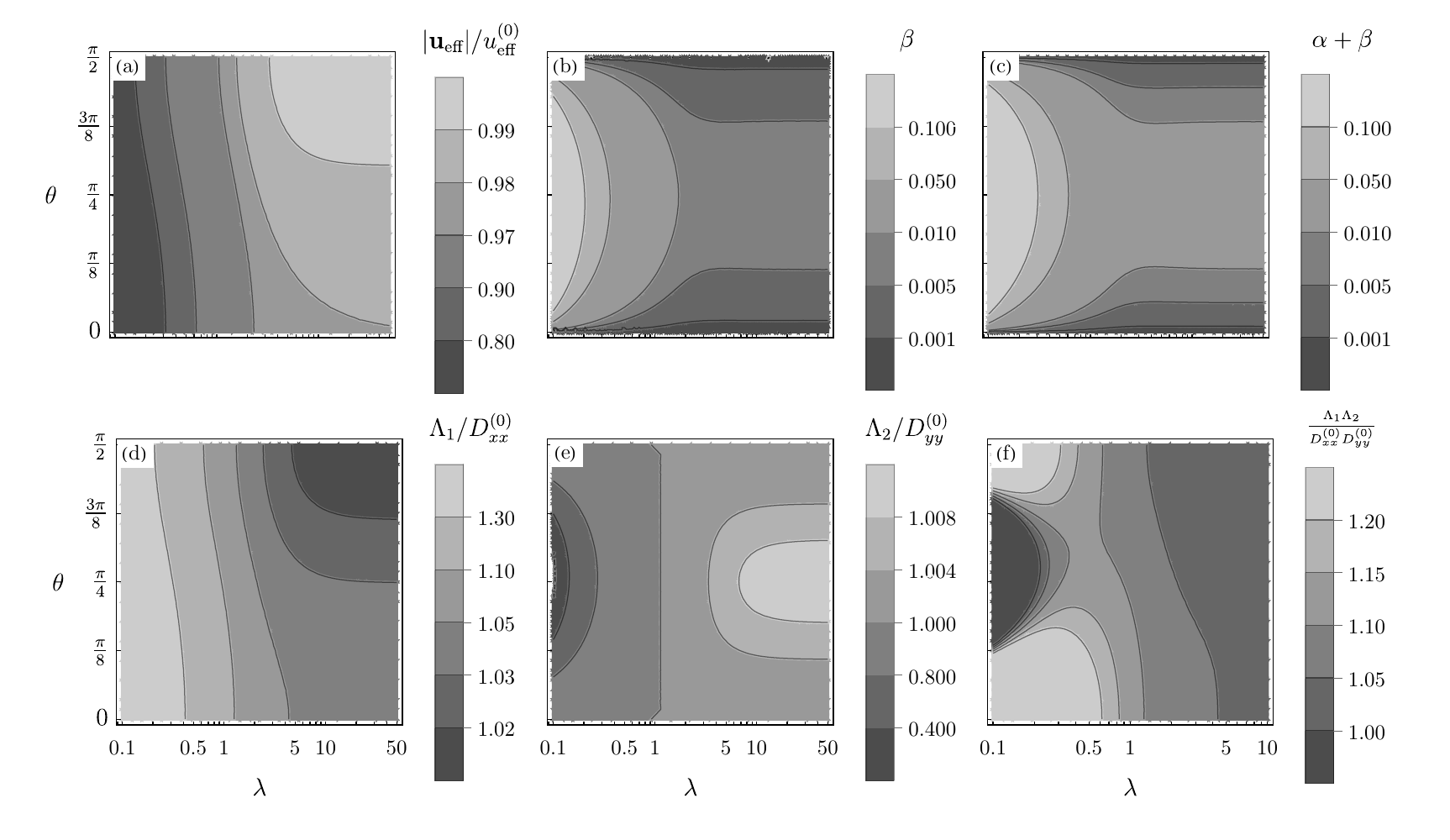}
    \caption{Effective transport coefficients corresponding to a single-mode herringbone surface structure, including (a) the velocity magnitude $|\vec{u}_{\rm eff}|$, (b) the angle $\beta$ between drift velocity and base flow direction~$\mathbf{e}_x$, (c) the angle $\alpha + \beta$ between the principle eigenvector of the dispersion matrix and the drift velocity, (d) the principle eigenvalue $\Lambda_1$, (e) the secondary eigenvalue $\Lambda_2$, and (f) the product of the eigenvalues $\Lambda_1\Lambda_2$, which gives a representation of the effective two-dimensional spreading of particles in the flow. The eigenvalues are normalized by the corresponding dispersion tensor component for standard Taylor dispersion, $D_{xx}^{(0)}$ and $D_{yy}^{(0)}$ (\ref{eq:diff0}), while the velocity is normalized by base flow, $u^{(0)}_{\rm eff}$ (\ref{eq:u0}).
    Results are shown for different wavelengths $\lambda$ and angles $\theta$ for $\cos(\theta) = \vec{k}\cdot \vec{e}_x/|\vec{k}|$, which is the angle between the wave vector~$\vec{k}$ and the base flow direction $\vec{e}_x$. Here, the P{\'e}clet number is Pe~$=100$ and the roughness is $\epsilon=0.1$.
    }
    \label{fig:singleMode}
\end{figure*}

Keeping $\theta=\pi/4$, a quantitative comparison of the effective transport parameters obtained from simulations across a range of $\lambda$ shows good agreement with our theoretical predictions. In particular, the results in \fref{fig:simulations}~(a,b) confirm that our theory very closely captures the effective drift of the particles in both directions. Note that for a surface with a wavelength of $\lambda/h\sim 1$ we expect that the particle experiences an effective drift in the $-\vec{e}_y$ direction with a velocity of about 2\% of the mean speed. However, the same surface can lead to reductions of almost 10\% in the direction of applied forcing.

In addition to the drift, the surface corrugations amplify the dispersion. \fref{fig:simulations}(c) shows the predicted values of $\Lambda_1$ and $\Lambda_2$ (corresponding to the eigenvalues of the dispersion matrix) normalized by the equivalent diffusion constant from classical Taylor dispersion theory (\ref{eq:diff0}). Note that we will maintain this normalization for the rest of this paper. We find that our simulation results match the theoretical predictions for $\Lambda_1$ well, with greater enhancement of diffusion at shorter wavelengths before settling to a constant value at longer wavelengths. Furthermore, the results show that $\Lambda_{1}\gsim D^{(0)}_{xx}$, which is indicative of the surface roughness continuing to have an effect on dispersion even when the wavelengths $\lambda$ of the disturbances are very large. On the contrary, our theory predicts that $\Lambda_2$ is decreased at small $\lambda$ and becomes slightly increased at large~$\lambda$. It is important to note, however, that at the P{\'e}clet number and wavelength $\lambda$ used here $\Lambda_2$ is comparable to molecular diffusion and hence the effect is so small that we could not reliably resolve it from our simulations.

Having verified our theoretical predictions with these simulations, in \fref{fig:singleMode} we study the effects of surface roughness across a wide range of surface corrugations with different $\lambda$ and $\theta$. We note that our theory is expected to be valid as the ratio $\epsilon/\lambda\to 0$~\cite{Kamrin:2010}, and therefore we expect that for small $\lambda$ our results may become amplified and detailed studies including the full hydrodynamic flows should be done in the future.

The surface structure tends to decrease the effective drift $|\vec{u}_{\rm eff}|$, although this effect gets smaller at higher wavelengths [\fref{fig:singleMode}(a)]. The magnitude of drift is only weakly dependent on $\theta$ at low wavelengths but increases with $\theta$ at high wavelengths. Longer wavelength surfaces with a wave vector oriented perpendicular to the base flow direction (grooves aligned with the flow direction) tend to have the smallest effect on dispersion, as any given tracer particle effectively experiences a locally flat surface.

The orientation  $\beta$ of the drift velocity relative to the forcing direction $\vec{e}_x$, is maximally reduced for $\theta = 0$ and $\pi/2$ and for long wavelengths, with the highest deflections predicted at low wavelengths for $\theta\approx \pi/4$ [\fref{fig:singleMode}(b)]. Note that this angle follows the sign convention depicted in \fref{fig:ellipses}, such that a positive surface orientation angle $\theta$ leads to drift in the negative $\vec{e}_y$ direction.

It is important to emphasize that the principal direction of dispersion does not align with the mean flow direction. In the absence of Brownian effects, as shown by Ref.~\cite{Chase:2022} for sedimenting spheres and Ref.~\cite{Kurzthaler:2022} for point particles  in shear flow, particles moving near a wavy surface exhibit helical trajectories due to hydrodynamic couplings with the surface structure, which leads to transport across the grooves. We expect that similar helical flow trajectories lead to the observed difference between the induced flow direction and the principal axes of diffusion. The variations in the angle between the principle eigenvector and the drift velocity, $\alpha + \beta$, is qualitatively similar to that of $\beta$ [\fref{fig:singleMode}(c)]. We note that the angle between the principle eigenvector and the base flow $\alpha$ is not depicted but behaves similarly, although with a smaller magnitude.

Turning now to the dispersion, small surface wavelength ($\lambda \sim 0.5$) generate more than 10\% increased dispersion along the principal eigenvector [\fref{fig:singleMode}~(d)]. This effect is more pronounced when the corrugations are perpendicular to the applied forcing ($\theta=0$) and decreased when they are aligned with the applied forcing ($\theta=\pi/2$). At large $\lambda$ the dispersion enhancement is reduced, as the perturbation flows driven by the surface topology weaken. Dispersion is observed to be enhanced only by a few percent at $\lambda = 50$. At shorter wavelengths, the dispersion is primarily controlled by changes in $\lambda$, while $\theta$ plays a secondary role. However, as $\lambda$ gets large, $\theta$ plays the dominant role in determining the strength of dispersion, with less enhancement as the wave vector approaches an angle of $\theta = \pi/2$ (corresponding to surface grooves aligned along $\vec{e}_x$).

The behavior of $\Lambda_2$ for a range of surfaces [\fref{fig:singleMode}(e)] is markedly different from that of $\Lambda_1$. For shorter wavelengths, dispersion along the secondary eigenvector is less than classical Taylor dispersion (and hence molecular diffusion). This reduction is strongest near $\theta=\pi/4$ and falls off symmetrically. For $\lambda\sim1$ the difference between the surfaces with roughness and classical theory is small. At large $\lambda \gsim 1$ the dispersion in this direction is enhanced relative to classical theory, with maximal enhancement for $\theta =\pi/4$. However, this effect is small, contributing less than 1\% to an increase in dispersion.

Finally, we consider the product of the two eigenvalues $\Lambda_1\Lambda_2$ [\fref{fig:singleMode}(f)], which is proportional to the area of the level curves of (\ref{eq:bivariate}) describing the concentration of the tracer particles. For short wavelengths, $\lambda\lsim 1$, the angle $\theta$ plays a large role in the dispersion of particles, with enhancements of over 20\% or reductions in spreading relative to classical Taylor dispersion depending on the orientation angle $\theta$. For longer wavelengths $\lambda \gsim  1$, the spreading is  maximized for $\theta=0$ and roughness can provide enhancements to the overall spreading of around $5\%$ for $\lambda\gsim 5$.

\subsection{Bumpy surface (two modes)}
Having analyzed several important features of single-mode surfaces, we now turn to bumpy surfaces consisting of two modes.  Here, we are particularly interested in symmetrical bumpy surfaces, such as that depicted in \fref{fig:surfaces}(b). In this case, the two wave vectors $\mathbf{k}_1$ and $\mathbf{k}_2$ are related by $k_{1,x} = k_{2,x}$ and $k_{1,y} = - k_{2,y}$ ($\vec{k}_1\cdot\vec{k}_2=0$). Furthermore, we take $H_1 = H_2 = 1/2$ in (\ref{eq:heightsbumps}) so that the maximum height of the bumps is always 1. We also describe the surfaces in terms of an angle $\theta$ measured relative to $\vec{k}_1$, such that $\theta = \arctan(k_{1,y}/k_{1,x})$, and a wavelength $\lambda = 2\pi/|\vec{k}_1|$.

Due to the reflectional symmetry of the wave vectors, the dispersion tensor for the resulting surface becomes diagonal when examined in the lab frame. Mathematically, the corrections to the dispersion tensor resulting from each surface mode are even in $k_y$ for the diagonal terms but odd in $k_y$ for the off-diagonal terms (see~\ref{appendix:difTen}). As shown in (\ref{eq:diffBumps}), the dispersion for each mode is additive and hence the corresponding off-diagonal contributions cancel when choosing wave vectors of this form.

Combining wave vectors in this manner results in a surface that does not have a preferred direction. Therefore any drift along $\vec{e}_y$ is eliminated and we can analyze the diffusion tensor in the lab frame, as the two non-zero components are the eigenvalues of the tensor. While the diffusion is still anisotropic, as diffusion in the $\vec{e}_x$ direction is still much greater than that in the $\vec{e}_y$ direction, the result is an elliptical distribution of tracers whose axes are aligned with the lab frame.
\begin{figure}[tp]
    \centering
    \includegraphics[width=\linewidth]{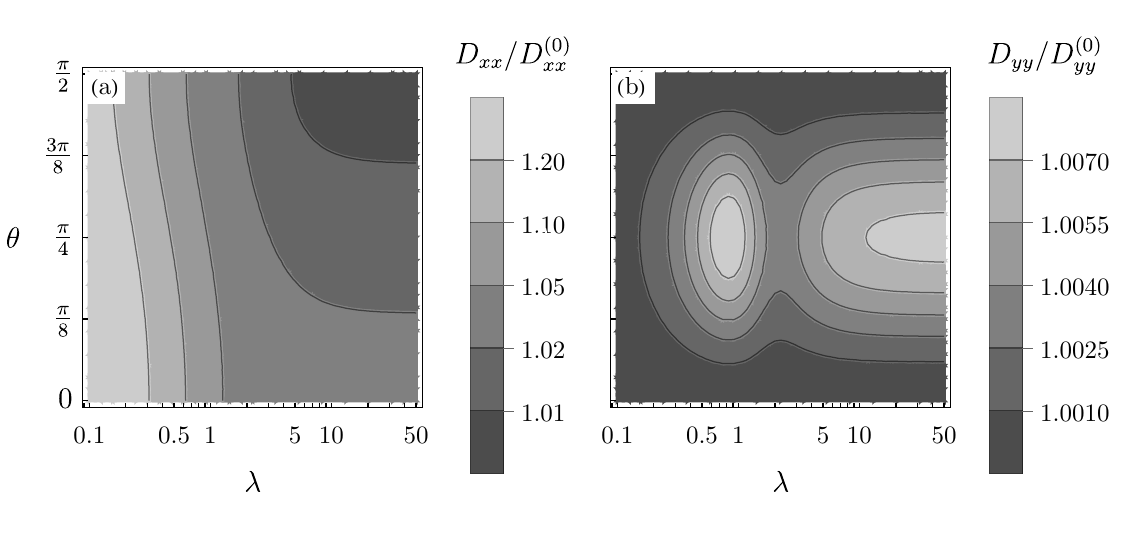}
    \caption{Components of the dispersion matrix, $D_{xx}$ and $D_{yy}$, for a bumpy surface consisting of two modes whose wave vectors $\mathbf{k}_1$ and $\mathbf{k}_2$ are related by $k_{1,x} = k_{2,x}$ and $k_{1,y} = - k_{2,y}$. Results are normalized by the predictions for classical Taylor dispersion, $D_{xx}^{(0)}$ and $D_{yy}^{(0)}$ (\ref{eq:diff0}). Further, the angle is defined by $\theta = \arctan(k_{1,y}/k_{1,x})$ and $\lambda=2\pi/|\vec{k}_1|$ denotes the wavelength. ere, the P{\'e}clet number is Pe~$=100$ and the roughness is $\epsilon=0.1$.
    }
    \label{fig:combinedBump}
\end{figure}
Examining $D_{xx}$, as displayed in \fref{fig:combinedBump}(a), we find that the bumpy case is very similar to the single-mode case, as the first component of the tensor is even with respect to $k_y$. Therefore, we are just adding mirror-symmetric components, which results in the same profile, with a weak effect of $\theta$ at short wavelengths, while at long wavelengths changes in $\theta$ dominate the behavior of $D_{xx}$. The largest increase to diffusion is observed for $\theta =0$, which results in an increase larger than 5\%  above classical Taylor dispersion for $\lambda =1$.

We observe a different behavior for $D_{yy}$ for the two-mode case than in the single-mode case, see \fref{fig:combinedBump}(b). There exists a local maximum of $D_{yy}$ for $\theta = \pi/4$ and $\lambda \sim 1$, which decreases for intermediate $\lambda$ before increasing again at large $\lambda$. It is important to note that these effects are still less than a 1\% increase relative to Taylor's classical result near a planar wall. However, the presence of this non-monotonic behavior allows one to potentially optimize cross-streamline dispersion of tracer particles by tuning the surface topography.

\subsection{Random surface ($N$ modes)}
Finally, we address the case of a randomly structured surface consisting of $N$ modes. We restrict ourselves to considering surfaces where the surface shape is described by
\begin{equation}
H(\vec{r}_\parallel) = \frac{1}{2N}\sum_{i,j=1}^N \gamma_{ij}\left( \cos(\vec{k}_{ij}\cdot\vec{r}_\parallel)+ \sin(\vec{k}'_{ij}\cdot\vec{r}_\parallel)\right), \label{eq:random_surface2}
\end{equation}
with $\vec{k}_{ij} = 2\pi [i,j,0]^{T}/(\lambda N)$, $\vec{k}'_{ij} = 2\pi [i,-j,0]^{\rm T}/(\lambda N)$, and coefficients $\gamma_{ij}$. To model a random surface the latter are drawn from a Gaussian distribution with mean $\langle \gamma_{ij}\rangle = 0$ and unit variance $\langle \gamma_{ij}\gamma_{nm}\rangle = \delta_{in}\delta_{jm}$, where the brackets indicate an average over many surface realizations (see \fref{fig:surfaces}~(c) for a realization of this surface). This description creates a surface consisting of a random distribution of symmetrical bumps similar to the two-mode case with no directional bias, which is what we would expect of a randomly chosen surface, such as from sanding lines or machine tool marks.
From our analysis above, we find that the overall effect of each random mode is additive in the diffusion tensor (\ref{eq:randsurfD}) with
coefficients $\alpha_{ij}=\beta_{ij}= \gamma_{ij}$.

\begin{figure}[tp]
    \centering
    \includegraphics[width=\linewidth]{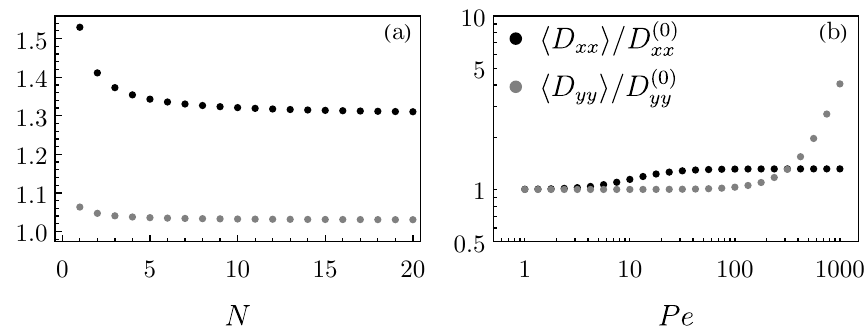}
    \caption{Average components of the dispersion matrix, $\langle D_{xx}\rangle$ and $\langle D_{yy}\rangle$, associated with a randomly bumpy surface (\ref{eq:random_surface2}) whose coefficients $\gamma_{ij}$ follow a normal distribution. Dispersion components (a) as a function of the number of random modes $N$ (with Pe~$= 100$, $\lambda=1$, and $\epsilon = 0.1$) and (b) as a function of the P{\'e}clet number (with $N=20$, $\lambda =1$, and $\epsilon = 0.1$). Results are normalized by the predictions for classical Taylor dispersion, $D_{xx}^{(0)}$ and $D_{yy}^{(0)}$ (\ref{eq:diff0}).}
    \label{fig:randomsurface}
\end{figure}
Our theory allows us to characterize the average dispersion coefficients, $\langle D_{xx}\rangle$ and $\langle D_{yy}\rangle$, of tracers suspended near random surfaces, by taking the average over many surface realizations. As the number of modes in the surface increases, the diffusion constants appear to approach a limit larger than the diffusivities in classical Taylor dispersion, see \fref{fig:randomsurface}(a). $\langle D_{xx}\rangle$ exhibits over 30\% increase in diffusivity due to the random surface while the increase of $\langle D_{yy}\rangle$ is around 5\%. These results clearly indicate that the introduction of a  random surface topology, even with relatively small amplitude, can significantly increase the dispersion of tracer particles advected near that  surface structure. The enhancement for large $N$ is to leading order limited by the contribution of the smallest wave vector: $\langle \vec{D}\rangle \lsim \vec{D}^{(0)}+\epsilon^2\vec{D}^{(2)}(\vec{k}_{NN})$.

We can also use this random surface formulation to examine the effect of varying P{\'e}clet number. As shown in \fref{fig:randomsurface}(b), which plots the average dispersion coefficients for random surfaces with $N=20$, both components of dispersion are close to that of classical Taylor dispersion at small Pe, indicating that classical Taylor dispersion is dominant relative to the effects of surface roughness. As Pe increases the normalized $\langle D_{xx}\rangle/D_{xx}^{(0)}$ increases slightly but remains bounded. This is due to the normalization factor from classical Taylor dispersion, which includes terms of both $\mathcal{O}(1)$ (corresponding to the effect of molecular diffusion) and $\mathcal{O}({\rm Pe}^2)$ (corresponding to the effect of advection). 
As the contribution from surface roughness scales quadratically with Pe, due to the fact that it also arises from advection, $\langle D_{xx}\rangle/D_{xx}^{(0)}$ remains bounded for all Pe and approaches a 30\% increase due to roughness.

In contrast, $\langle D_{yy}\rangle/D_{yy}^{(0)}$ increases for ${\rm Pe}\gsim\mathcal{O}(1/\epsilon^2)$. Mathematically, this is due to the fact that  $D_{yy}^{(0)}$ only  depends on molecular diffusion while the advection contribution of $D_{yy}^{(2)}$ scales with ${\rm Pe}^2$. As the P{\'e}clet number increases advective contributions become more important than molecular diffusion and so the dispersion driven by this advection becomes more pronounced than the effects of molecular diffusion alone.

\section{Conclusion}
The impact of surface properties on the classical Taylor dispersion phenomenon~\cite{Taylor:1953, Aris:1956} in narrow tubes has been subject of several works~\cite{Hoagland:1985,Ng:2011, Mangeat:2017,Marbach:2019}. Here, we consider wide rectangular channels, which are confined along the vertical direction only, under shear flow and study the impact of structured surface topographies on the transport behavior of diffusing tracer particles. By generalizing Taylor's dispersion theory~\cite{Taylor:1953}, we derive analytical predictions for the probability distribution of tracer particles at coarse-grained length and time scales, in the limit of small surface roughness.

Our theory, valid for all surfaces expandable in terms of a Fourier series, shows that surface structures can induce asymmetric dispersion along a principal direction that is different from the main flow direction. This effect is prominent for simple surface corrugations. For a P{\'e}clet number of 100, these corrugations can lead to enhancements of 20\% along the principle direction while reducing the dispersion in the perpendicular direction. This can generate enhanced particle spreading relative to classical Taylor dispersion theory. Furthermore, our results reveal that surface corrugations that are tilted with respect to the forcing direction induce  drift of tracer particles along the corrugations and reduce the drift along the forcing direction. Tracer particle drift along corrugations has been observed in experiments of non-Brownian particles in shear flow~\cite{Choi:2007, Choi:2011,Hsu:2008} and under gravity~\cite{Chase:2022}. Our study, however, provides the first quantitative treatment of Brownian effects. We further show that for surfaces that do not have a directional bias the dispersion tensor becomes diagonal within the lab frame.

In addition, we prove that the contributions of individual wave modes of the surface to the large-scale diffusivity and drift velocity can be decoupled to leading order in the surface roughness. This allows us to consider randomly structured surfaces by summing over the contributions of each wave number. Within the limits of our theory, we find that randomly structured surfaces lacking a directional bias always enhance dispersion in both directions due to resulting disturbance flows. Under some conditions, we have observed that relative enhancement can reach an asymptotic value around 30\% in the direction of the forcing as the number of random modes grows; this effect varies slightly with the P{\'e}clet number as demonstrated in \fref{fig:randomsurface}(b).

Our predictions, irrespective of the details of the surface, show that the impact of surface structure for motion along the main flow direction is always orders of magnitude larger than for motion perpendicular to it. The latter is always an effect coming from the (weak) perturbation flow generated by the surface roughness. Thus, this particular feature comes from our assumption of small surface roughness and is expected to become  amplified for surface roughness of the order of the channel width, as in porous media or membranes, where restrictions can reduce transport along the flow direction. It would be interesting to study the transition from flow-enhanced dispersion to a potential slow-down due to entropic constraints in this context.
Our results further indicate that small amplitude surface roughness can slow down tracer transport across the entire width of a channel while maintaining the same flow conditions, which may be useful for applications, such as ensuring a chemical reaction to complete.

\ack J.V.R. and H.A.S. thank the NSF for support via CBET-2127563 and MCB-1853602. C.K. acknowledges support from the Austrian Science Fund (FWF) via the Erwin Schr{\"o}dinger fellowship (Grant No. J4321-N27). 
\pagebreak
\section{Appendix}
\appendix
\section{Roughness-induced velocity fields \label{appendix:flow}}
\subsection{3D wavy surface: one mode}
Using  $\vec{k}=k[\cos\theta, \sin\theta,0]^{\rm T}$ the roughness-induced velocities of first order assume the form:
\numparts
\begin{eqnarray}
    \fl
    \hat{u}^{(1)}(z; \vec{k})\equiv \frac{\rmd^2Q_1}{\rmd z^2}= -\frac{{\rm csch}(k)}{2\left(2 k^2-\cosh (2 k)+1\right)}\Big(k \big(\cos ^2(\theta ) (2 k \sinh (k (z+1))\nonumber\\ +z \cosh (k (z-3))  -2(z-1) \cosh (k (z-1)) +(z-2) \cosh (k (z+1))\nonumber\\-4 k z \sinh (k) \cosh (k z))  -k (\cos (2 \theta)-3) \sinh (k-k z)\big) \sinh (k (z-3))\nonumber\\+\sinh (k (z+1))+2 \sinh (k-k z)\Big)
\end{eqnarray}
\begin{eqnarray}
\fl \hat{v}^{(1)}(z; \vec{k})\equiv\frac{\rmd^2Q_2}{\rmd z^2}=\frac{k \sin (\theta ) \cos (\theta )}{2\left(2 k^2-\cosh (2 k)+1\right)} \Big(-z (4 k+\coth (k)) \cosh (k z)\nonumber \\+z {\rm csch}(k) \cosh(k (z-3))+\sinh (k z) (4 k \coth (k)+3 z-4)\Big)
    \end{eqnarray}
\begin{equation}
\fl
    \hat{w}^{(1)}(z;\vec{k}) = \frac{k \cos (\theta ) (2 k (z-1) \sinh (k z)+z \cosh (k (z-2))-z \cosh (k z))}{2 k^2-\cosh (2 k)+1}
\end{equation}
\endnumparts
The non-periodic second-order contributions evaluate to
\numparts
\begin{eqnarray}
    \fl \bar{u}_0^{(2)}(\vec{k}) = \frac{k \left(\sinh (2 k) (\cos (2 \theta)+3)-4 k \left(k \coth (k) \sin ^2(\theta)+2 \cos ^2(\theta)\right)\right)}{8 k^2-4 \cosh (2 k)+4}, \\
    \fl \bar{v}_0^{(2)}(\vec{k}) = \frac{k \sin (2 \theta) (\sinh (2 k)+2 k (k \coth (k)-2))}{8 k^2-4 \cosh (2 k)+4},\\
    \fl \bar{w}_0^{(2)}(\vec{k}) =0
\end{eqnarray}
\endnumparts
\subsection{General, randomly structured surface}
By linearity the first-order velocities induced by a surface of shape as in (\ref{eq:random_surface}) assume the form
\begin{eqnarray}
\fl
    \vec{u}^{(1)}(\vec{r}; \lambda, \{\alpha_{ij},\beta_{ij}\}) = \frac{1}{N} \sum_{i,j=1}^N\left(\alpha_{ij}\cos(\vec{k}_{ij}\cdot\vec{r}_\parallel)+\beta_{ij}\sin(\vec{k}_{ij}\cdot \vec{r}_\parallel)\right)\vec{Q}_{zz}(z; \vec{k}_{ij})\nonumber\\+\frac{1}{N} \sum_{i,j=1}^N\left(\alpha_{ij}\sin(\vec{k}_{ij}\cdot\vec{r}_\parallel)-\beta_{ij}\cos(\vec{k}_{ij}\cdot \vec{r}_\parallel)\right)\vec{k}_{ij}\cdot\vec{Q}_{z}(z;\vec{k}_{ij})\vec{e}_z.
\end{eqnarray}
The boundary conditions at $z=0$ for the second-order velocity field are
\begin{equation}
\fl
    \vec{u}^{(2)}(\vec{r}; \lambda, \{\alpha_{ij},\beta_{ij}\})\Bigl|_{z=0} = -\frac{1}{2N^2}\sum_{i,j=1}^N\left((\alpha_{ij})^2+ (\beta_{ij})^2\right) \vec{Q}_{zzz}(z=0;\vec{k}_{ij}) + {\rm p.t.},
\end{equation}
where we observe that the coupling of different surface modes only contributes to the periodic terms `p.t.'. The second-order velocities are of the form as in (\ref{eq:vel2_bumps}).
\section{Full form of the dispersion tensor} \label{appendix:difTen}
Noting that the dispersion matrix is symmetrical ($D_{12}=D_{21}$), we may express the analytic form of leading order correction to the dispersion matrix for a given wave number $\vec{k} = (k\cos(\theta),k\sin(\theta))^{{\rm T}}$ as:
\begin{eqnarray}
\fl
    D^{(2)}_{11} = -\frac{1}{1920 k^4 \left(-2 k^2+\cosh (2 k)-1\right)^2}\Bigg[1440 k^6 \sin ^4(\theta ) {\rm csch}^2\left(\frac{k}{2}\right)\nonumber
    \\ +k \sinh (4 k) \big(4 \left(k^4+105\right) \cos (2 \theta )
     -3 \left(55 \cos (4 \theta )-4 k^4+45\right)\big)\nonumber
     \\
     +960 \left(4 k^2-5\right) k \sin ^2(2 \theta ) \sinh (k)-160 \sin ^2(\theta ) \cosh (3 k) \big(\left(k^2-36\right) \cos (2 \theta )\nonumber
    \\
    -k^2-12\big)-80 \sin ^2(\theta ) \cosh (4 k) \left(\left(k^2+18\right) \cos (2 \theta )-k^2+6\right)\nonumber
    \\+64 \left(2 k^2+1\right) k^6 \cos ^2(\theta )
    +32 \left(k^4-90\right) k^5 \sin ^2(\theta ) \coth \left(\frac{k}{2}\right)\nonumber
    \\-320 k^2\cos ^2(\theta ) \left(\left(2 k^4+22 k^2+15\right)\cos (2 \theta )-2\left(k^4+11 k^2+3\right)\right)\nonumber
    \\
    +80 k^2 \left(k^2 \left(\left(2 k^2-7\right) \cos (4 \theta )-3 \left(2k^2+9\right)\right)+2 \left(-2 k^4+7 k^2+6\right) \cos (2 \theta )\right)\nonumber
    \\-\frac{320 \left(5 k^2+48\right) k^4 \sin ^4(\theta )}{\cosh (k)+1}
    +160 \sin ^2(\theta ) \cosh (k) \Big(-8k^4-k^2 \nonumber
    \\+\left(8 k^4-191 k^2-36\right) \cos (2 \theta )-12\Big)+40 \Big(8 k^6+37 k^4-123 k^2\nonumber
    \\+4 \left(-4 k^6-9 k^4+85 k^2+15\right) \cos (2 \theta )+\left(8 k^6+23 k^4-133k^2-45\right) \cos (4 \theta ) \nonumber
    \\
    -15\Big)-2 k \sinh (2 k) \big(32 k^6+72 k^4-1350 k^2+12 \left(7 k^4+90 k^2+35\right) \cos (2 \theta )\nonumber
    \\+5 \left(4 k^4-90 k^2+159\right) \cos (4 \theta)-1095\big)-4 \cosh (2 k) \big(8 k^6+120 k^4 \nonumber
    \\
    +645 k^2+5 \left(16 k^4+59 k^2-72\right) \cos (4 \theta )+4 \Big(2 k^6-30 k^4-145 k^2\nonumber
    \\+120\Big) \cos (2 \theta )
    -120\big)+32 k^3\sin ^2(\theta ) \tanh \left(\frac{k}{2}\right) \Big(-480 \cos (2 \theta )+k^6 \nonumber
    \\-90 k^2-480\Big)
    -960 k \sin ^2(2 \theta ) \sinh (3 k)\Bigg]
\end{eqnarray}
\begin{eqnarray}
    \fl D^{(2)}_{12} = \frac{1}{7680 k^4 \left(-2 k^2+\cosh (2 k)-1\right)^2}\Bigg[{\rm csch}^2(k) \Bigg(2 \sin (2 \theta ) \Big(44 k^5 \sinh (4 k)-k^5 \sinh (6 k)\nonumber
    \\
    -40 \left(k^2-12\right) \left(8 k^2+3\right) \cosh (3 k)
    +40 \left(k^2-12\right) \cosh (5 k)+20\Big(k^2 \nonumber
    \\+6\Big) \cosh (6 k)+\left(16 k^4-1525\right) k^5 \sinh (2 k)+4 \big(2 k^6-30 k^4-155 k^2 \nonumber
    \\
    +60\big) \cosh (4 k)+80 \Big(54 k^3 \sinh ^3(k)+\left(k^2-12\right) \left(8k^4+4 k^2+1\right) \nonumber \\
    -42 k \sinh ^5(k)\Big) \cosh (k)+4 \big(8 k^8+446 k^6
    +1850 k^4+1495 k^2+420\big)\nonumber
    \\
    -4 \left(8 k^8-112 k^6-100 k^4+1345 k^2+510\right) \cosh (2 k)\Big)\nonumber
    \\-5 \sin (4\theta ) \Big(k \Big(4 k \Big(\big(378
    -16 k^2\big) \cosh (3 k)-61 \cosh (4 k)+2 \cosh (5 k) \nonumber
    \\
    +\cosh (6 k)+k \Big(-480 \sinh (k)
    +294 \sinh (2 k)-96 \sinh (3 k)+45 \sinh (4 k)\nonumber
    \\ -16 k\Big(\cosh (4 k)
    +k \sinh ^3(k)\cosh(k)\Big)\Big)\Big)\nonumber
    \\ -768 \sinh ^5\left(\frac{k}{2}\right) \cosh ^3\left(\frac{k}{2}\right) (-10 \cosh (k)+11 \cosh (2 k)+43)\Big)\nonumber
    \\+16\Big(8 k^6-92 k^4 -95 k^2-36\Big) \cosh (k)
    +4 \Big(32 k^6-108 k^4\nonumber
    \\ -533 k^2-306\Big) \cosh (2 k)+4 \big(80 k^6+508 k^4+593 k^2+216 \cosh (3 k)\nonumber\\+36 \cosh (4 k)-72 \cosh (5 k)+18\cosh (6 k)+252\big)\Big)\Bigg)\Bigg]
\end{eqnarray}
\begin{eqnarray}
    D^{(2)}_{22} =  \frac{1}{192 k^4 \left(-2 k^2+\cosh (2k)-1\right)^2} \Bigg[\sin ^2(2 \theta ) {\rm csch}^2(k) \Big(-80 k^6-508 k^4\nonumber
    \\ -593 k^2+k \Big(k \big(2 \left(8 k^2-189\right) \cosh (3 k)
    +61 \cosh (4 k)-2 \cosh (5 k)\nonumber
    \\
    -\cosh (6 k)+k \big(480 \sinh(k)-294 \sinh (2 k)+96 \sinh (3 k)-45 \sinh (4 k) \nonumber
    \\
    +16 k \left(\cosh (4 k)+k \sinh^3(k) \cosh(k)\right)\big)\big) \nonumber
    \\
    +192 \sinh ^5\left(\frac{k}{2}\right) \cosh^3\left(\frac{k}{2}\right) (-10 \cosh (k)+11 \cosh (2 k)+43)\Big)\nonumber
    \\
    +4 \left(-8 k^6+92 k^4+95 k^2+36\right) \cosh (k)+\Big(-32 k^6+108 k^4+533 k^2\nonumber
    \\
    +306\Big) \cosh (2 k)-216 \cosh (3k)
    -36 \cosh (4 k)+72 \cosh (5 k)\nonumber
    \\-18 \cosh (6 k)-252\Big)\Bigg]
\end{eqnarray}
\endnumparts
\section{Brownian dynamics simulations \label{appendix:BD}}
To validate our analytical predictions for the effective drift and diffusivities, we perform computer simulations of Brownian particles in shear flow near corrugated surfaces. Therefore, we assume that the particles are neutrally buoyant and small compared to the surface corrugations, so that they can be approximated by point particles, which, in addition to diffusion, follow the flow stream lines. The equation of motion reads
\begin{eqnarray}
    \frac{\rmd \vec{r}}{\rmd t} &= \vec{u}+\sqrt{2D_0}\boldsymbol{\xi}(t),
\end{eqnarray}
where we have modeled Brownian motion
in terms of an independent Gaussian white noise process
$\boldsymbol{\xi}(t)$ of zero mean and delta correlated
variance $\langle \xi_{i}(t)\xi_{j}(t')\rangle = \delta_{ij}\delta(t-t')$ with $i,j = 1,2,3$. We use the flow field $\vec{u}$ from our perturbation approach (as outlined earlier) and employ reflective boundary conditions at the walls. Further, we non-dimensionalize via
\begin{eqnarray}
    \vec{r}&= h\hat{\vec{r}} \qquad t= \frac{h^2}{D_0} \hat{t} \qquad \vec{u}= u_0\hat{\vec{u}},
\end{eqnarray}
where $\hat{\vec{r}}$, $\hat{t}$, and $\hat{\vec{u}}$ are dimensionless quantities, and discretize the equation of motion
\begin{eqnarray}
\hat{\vec{r}}(\hat{t}+\Delta\hat{t})&= \hat{\vec{r}}(\hat{t})+ {\rm Pe}\, \hat{\vec{u}}\Delta\hat{t} +\sqrt{2 \Delta \hat{t}}\ \vec{N}_\xi.
\end{eqnarray}
Here, $\Delta \hat{t}$ denotes the discretized time step and $\vec{N}_\xi$ is a vector with independent and normally distributed random variables with zero mean and unit variance. To corroborate our theoretical predictions with simulations, we use a P{\'e}clet number of Pe~$=100$, choose $H(\vec{r}_\parallel)=\cos(\vec{k}\cdot\vec{r}_\parallel)$ with $\vec{k}=(2\pi/\lambda)[\cos\theta,\sin\theta]^{\rm T}$ and $\theta=\pi/4$, and vary the wavelength $\lambda\in[0.5,10]/\sqrt{2}$. We simulate $5\cdot 10^3$ trajectories using Python and use a time-step of $\Delta\hat{t}=10^{-4}$.

We then compute the average displacements along $\langle \Delta x\rangle$ and $\langle \Delta y\rangle$ to extract the drift.  Subtracting the drift from the displacements and realigning the displacements along the main direction of diffusion, we compute the mean-square displacements and extract the long-time diffusivities along the main direction of diffusion (and hence the dispersion matrix).
\pagebreak
\bibliography{literature.bib}
\end{document}